\documentclass[twocolumn,aps,prl,10pt]{revtex4-1}

\usepackage{amsmath}
\usepackage{amsfonts}
\usepackage{color}
\usepackage{graphicx}
\usepackage{float}
\usepackage{cancel}
\usepackage{xcolor}
\usepackage{chngcntr}

\begin{document}

\title{Supersolid edge and bulk phases of a dipolar quantum gas in a box}

\author{S. M. Roccuzzo}
\author{S. Stringari}
\author{A. Recati}
\affiliation{INO-CNR BEC Center and Dipartimento di Fisica, Universit\`a degli Studi di Trento, 38123 Povo, Italy}
\affiliation{Trento  Institute  for  Fundamental  Physics  and  Applications,  INFN,  38123,  Trento,  Italy}

\date{\today}

\begin{abstract}
 
 We investigate the novel density distributions acquired by a dipolar Bose-Einstein condensed gas
    confined in a box potential, with special focus on the effects of supersolidity.  Differently from the case of harmonic trapping, the ground state
    density reveals a strong depletion in the bulk region and an accumulation of atoms near the walls, well separated from the bulk,
    as a consequence of the competition between the attractive and the  repulsive nature of the  dipolar force. In a quasi
    two-dimensional geometry characterized by cylindrical box  trapping,   we observe the emergence of a ring-like configuration near the boundary of the box, reveling peculiar supersolid and crystal effects in a useful range of parameters.  In the case  of   square box trapping   the density oscillations along the edges, caused by the enhanced accumulation of atoms near the vertices,  exhibit  interesting analogies  with  the case of box trapped one dimensional configurations.
    For sufficiently large  values of the atom number also the bulk region can exhibit supersolidity, the resulting  geometry  reflecting  the symmetry of the
    confining potential even for large systems.
    
\end{abstract}

\maketitle
\paragraph{Introduction.} Bose-Einstein condensed atomic gases have proved to be an invaluable tool for the study of the
physics of many body systems. However, while typical many body problems consider translationally
invariant systems in the thermodynamic limit, Bose-Einstein condensates (BECs) are ordinarily
realized in small, inhomogeneous samples confined by harmonic potentials \cite{BecBook2016}.
Although harmonic trapping allows the study of relevant properties of these many body systems (e.g.
collective excitations \cite{stringari96,mewes96}, superfluid properties
\cite{guery_odelin1999,marago2000,Rossi2016}, quantized vortices
\cite{madison2000,AboShaeer2001,Haljan2001}), other important properties, like sound propagation or
critical behaviors, can be better studied in uniform systems. For these reasons, Bose-Einstein
condensation in ``box'' potentials has been an emerging topic of research in recent years, leading
to the realization of uniform BECs in gases of alkali atoms and first important measurements in both
3D and 2D
configurations\cite{Raizen2005,Hadzibabic2013,Gupta2005,Navon167,Chomaz2015,Navon2016,Ville2017,Ville2018}.

The achievement of BECs of magnetic atoms in harmonic traps \cite{BEC_Cr,BEC_Dy162,BEC_Dy164,BEC_Er} opened the way to the study of the very
peculiar phenomena, which includes a geometry dependence of the phase diagram
stability \cite{Koch2008}, a rotonized excitation spectrum
\cite{SantosRoton,Roccuzzo1,Chomaz2018,Petter2019}, quantum droplets
\cite{ferrier_barbut2016,Kadau2016,ferrier_farbut_2016_2,Schmitt2016} and more recently supersolidity
\cite{F1,roccuzzo2020,I1,I2,S1,S2}.
While most of the theories for supersolidity are developed for
infinite systems all the
experiments have been so far conducted in harmonic traps.

\begin{figure}
    \centering
    \includegraphics[width=0.5\textwidth]{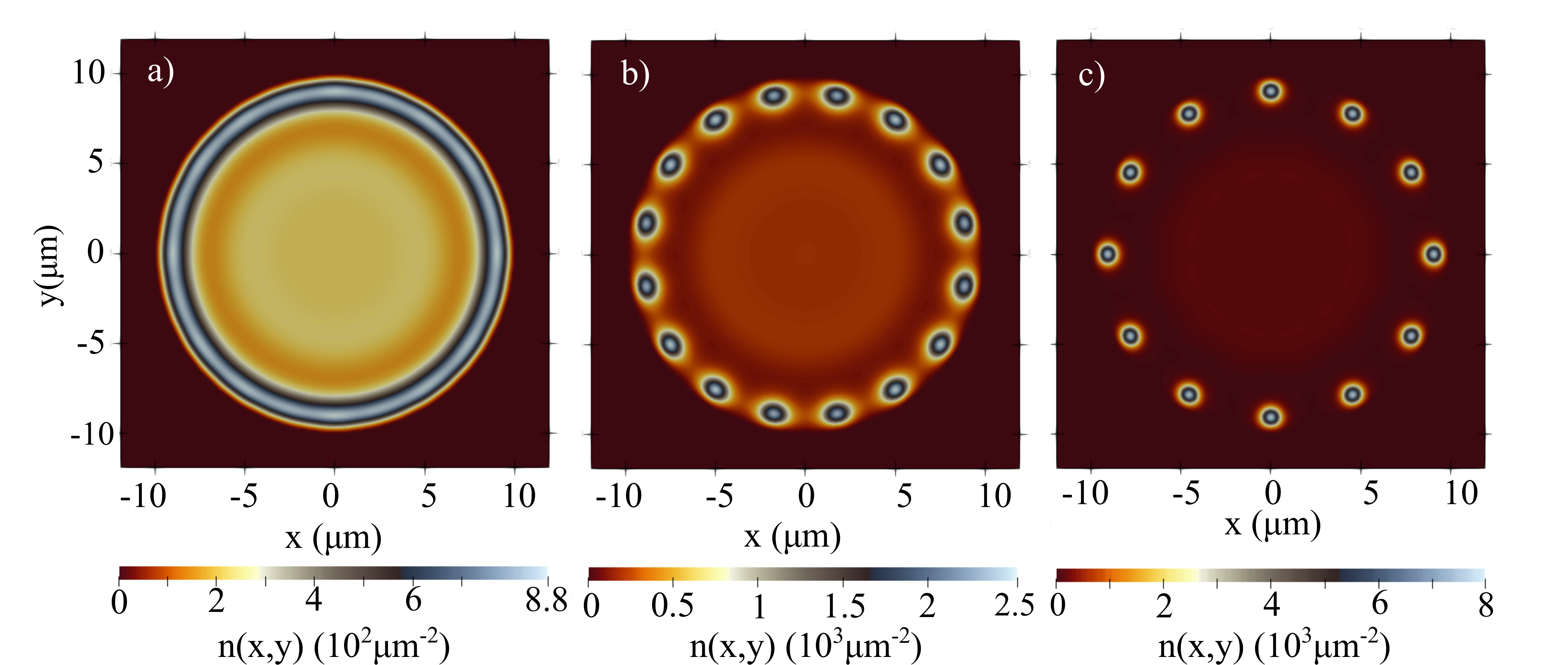}
    \caption{Ground state integrated density profiles $\mbox{n(x,y)=}\int\mbox{dz}|\Psi\mbox{(x,y,z)}|^2$ 
    for a gas of $10^5$ atoms of $^{164}$Dy confined in the polarization direction by a harmonic potential of frequency $\omega_z=(2\pi)100$Hz, and by a box potential in the x-y plain, with the shape of a circle of radius R=$10.185\mu$m. The value of $\epsilon_{dd}$ for panels a), b) and c) is, respectively, 1.32, 1.404 and 1.467. The height of the box is fixed to $\mbox{V}_0=100\hbar\omega_z$}
    \label{circular_box}
\end{figure}

\begin{figure}
    \includegraphics[width=0.5\textwidth]{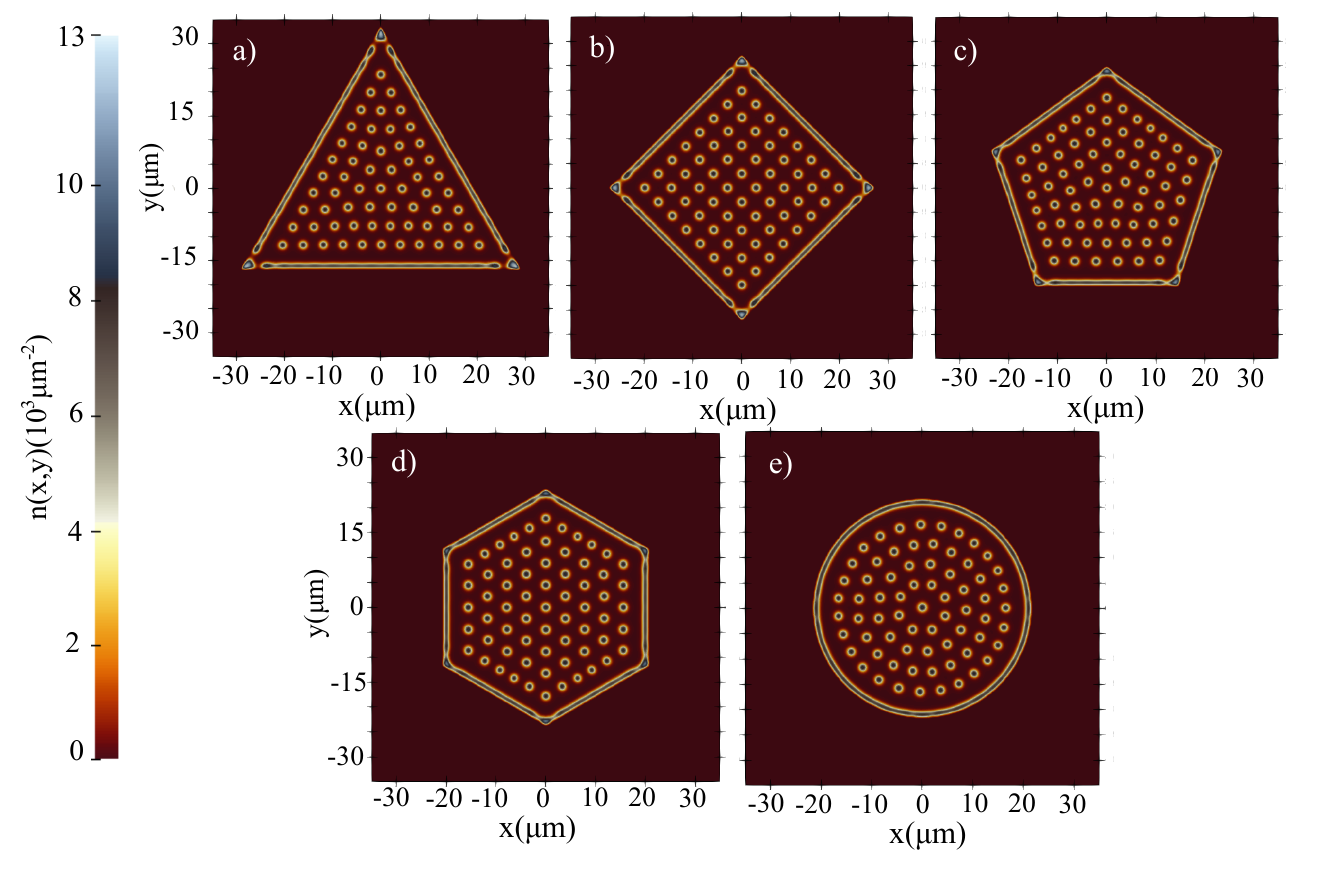}
    \caption{Ground state integrated density profiles $\mbox{n(x,y)=}\int\mbox{dz}|\Psi\mbox{(x,y,z)}|^2$ for a gas of $2\times10^6$ atoms of $^{164}$Dy confined in the polarization direction by a harmonic potential of frequency
    $\omega_z=(2\pi)100$Hz, and by a box potential in the x-y plain with the shape of a triangle of side $L=58.58\mu m$ (panel a), a square of side $L=45.84\mu m$ (panel b), a pentagon of side $L=38.33 \mu m$ (panel c), an hexagon of side $L=33.82 \mu m$ (panel d), and a circle of radius $R=21.75\mu m$ (panel e). The value of $\epsilon_{dd}=1.36$ is the same for all the configurations, which also have the same area.}
    \label{geometries}
\end{figure}

The natural question which arises is therefore how a dipolar gas behaves in a box potential, and to
what extent its configurations mimic their thermodynamic counterparts. First theoretical investigations carried out in the  deep superfluid phase \cite{Lu2010}  have pointed out the peculiar phenomenon of accumulation of the density distribution near the boundary, as a consequence of the repulsive behavior of the aligned dipoles. This effect is  strongly reduced  in the presence of transverse harmonic trapping because of the  high energetic cost for dipoles to move away from the center of the trap. The present work is based on a proper generalization of the Gross-Pitaevskii equation which contains the terms due to the Lee-Huang-Yang (LHY) correction to the mean field equation of state and it mainly focuses on the new supersolid features exhibited by the system  in the presence of the box. We find that in two-dimensional geometries the accumulation along the border is enhanced in the regimes where the LHY correction is relevant, creating edges pretty well separated from the bulk. For a relatively small number of atoms, the bulk remains in a low density superfluid phase, while the edges can show typical supersolid or droplet crystal structures (see Fig. \ref{circular_box}). Increasing the atom density leads to a supersolid bulk region, while the edges can be found to be in a high-density superfluid phase (see Fig. \ref{geometries}). Moreover, the lattice emerging in the bulk has not in general a triangular (or honeycomb) pattern, as expected for an infinite system \cite{Pohl2019}, but its structure is dictated by the shape of the confining box potential even for relatively large systems.

\paragraph{The Model.} We start our exploration by considering  the case of a quasi 2-dimensional dipolar BEC, obtained by imposing a harmonic
confinement only in the polarization direction ($z$-axis). 
At zero temperature, the dipolar BEC is described by a macroscopic wave function $ \Psi({\bf r},t) $, whose square modulus gives the
local density of the system, and which obeys to the so-called extended Gross-Pitaevskii equation
(eGPE) \cite{Pelster,Wachtler2016}

\begin{align}
    &i\hbar\frac{\partial \Psi({\bf r},t)}{\partial t} = 
     \Bigl[-\frac{\hbar^2\nabla^2}{2m}+V_{ext}({\bf r}) + g|\Psi({\bf r},t)|^2 \nonumber \\
    & + \int d{\bf r}' V_{dd}({\bf r}-{\bf r}')|\Psi({\bf r}',t)|^2
 + \gamma(\epsilon_{dd})|\Psi({\bf r},t)|^3 \Bigr] \Psi({\bf r},t)
 \label{eGPE}
\end{align}
where $V_{ext}({\bf r})$ is the trapping potential, $g=4\pi\hbar^2a/m$ is the coupling constant
fixed by the s-wave scattering length $a$, $V_{dd}({\bf r}) =
\frac{\mu_0\mu^2}{4\pi}\frac{1-3\cos^2\theta}{|{\bf r}|^3}$ is the dipole-dipole interaction between
two identical magnetic dipoles $\mu$ aligned along the z-axis ($\theta$ is the angle between ${\bf
r}$ and the z axis). A key dimensionless parameter is $\epsilon_{dd}$, defined as 
\begin{equation}
\epsilon_{dd}=\frac{\mu_0\mu^2}{3g}
\end{equation}
which measures the relative strength of the dipolar and the contact interaction.  Experimentally, the parameter $\epsilon_{dd}$ is tuned by changing the value of the $s$-wave scattering length $a$, thanks to the occurrence of a Feshbach resonance. The last term in the eGPE equation is the local density
approximation of the Lee-Huang-Yang correction to the ground state energy of the system
\cite{Pelster,Wachtler2016}, with 
\begin{equation}
\gamma(\epsilon_{dd})=\frac{16}{3\sqrt{\pi}}ga^{\frac{3}{2}}{\mathrm Re}[\int_0^\pi d\theta \sin(\theta)
[1+\epsilon_{dd}(3\cos^2\theta-1)]^{\frac{5}{2}}.
\end{equation}
The eGPE Eq. \ref{eGPE} provides a reliable description of the available experimental phenomenology.  In the absence of confinement in the transverse direction, this model predicts, that for a certain value of the density and of $\epsilon_{dd}$, a phase transition between a uniform superfluid and a supersolid occurs. In the thermodynamic limit, the supersolid lattice is predicted to be triangular or honeycomb \cite{Pohl2019}. The occurrence of such lattice symmetry has been also predicted for the case of transverse, radially symmetric, harmonic trapping \cite{gallemi2020,roccuzzo2020}. Very recently, the possible existence of other exotic configurations in harmonic traps has been proposed \cite{hertkorn2021,zhang2021}.   

\paragraph{Circular box potential.} We first consider the case where the  transverse confinement in the $x$-$y$  plane  is provided by a circular box, while the confinement along the polarization direction ($z$-axis) is of harmonic nature. The case of a square box in the $x$-$y$  plane will be discussed later, while the properties of a one-dimensional box will be discussed in the Supplementary Materials. We always fix the height of the box potential to a value large enough to ensure that the density goes practically to zero at the border. Similar configurations have been already experimentally realized to trap alkali atoms \cite{Navon167,Navon2016,Chomaz2015}. In Fig. \ref{circular_box} we show examples of the ground state density profiles (obtained by propagating equation \ref{eGPE} in imaginary time) of $N=10^5$ atoms of $^{164}$Dy, for different values of $\epsilon_{dd}$. As already anticipated, most of the atoms accumulate at the edge of the confining potential, forming a  quasi-one-dimensional ring structure well separated  from the atoms in the bulk. For small values of $\epsilon_{dd}$ both the edge and the bulk remain in the superfluid phase, while increasing $\epsilon_{dd}$ (i.e. increasing the effect of the dipolar force), the edge region clearly undergoes a phase transition to the supersolid phase, where the density peaks  near the boundary of the box  exhibit a finite overlap, ensuring global phase coherence. The overlap between the density peaks disappers for even larger values of $\epsilon_{dd}$,  the system forming a sort of one dimensional ring crystal. 

The emergent edge ring geometry allows to estimate the superfluid density in terms of the Leggett variational expression \cite{Leggett1970,Leggett1998}.  To  this purpose  we write the ground-state density in cylindrical coordinates $\rho(r,\theta,z)$, so that the Leggett's estimate for the superfluid density can be written as
\begin{equation}
  \frac{n_S}{n}=\frac{2\pi}{n} \left(\int \frac{d\theta}{\int dr dz \rho(r,\theta,z)}\right)^{-1}
  \label{leggett_supfrac}
\end{equation}
 where the integration over the radial coordinate is performed only in the edge region, identified by the density minima that appears both at the border of the box (where the density goes to zero), and at the interface between the edge and the bulk.  In the case of the  (extended) Gross-Pitaevskii equation the  estimate (\ref{leggett_supfrac}) for the superfluid density has been already verified to coincide with the result obtained by  imposing the twisted boundary condition to  a one-dimensional  array of droplets  \footnote{S. Roccuzzo, PhD Thesis, in preparation}. Remarkably in \cite{Martone2021} it has  been show that Leggett's estimate coincides with the exact superfluid density also in the case of peculiar   stationary non-ground state solutions (cnoidal wave solution) which exhibit a periodic density modulation. 
\begin{figure}
    \centering
    \includegraphics[width=0.4 \textwidth]{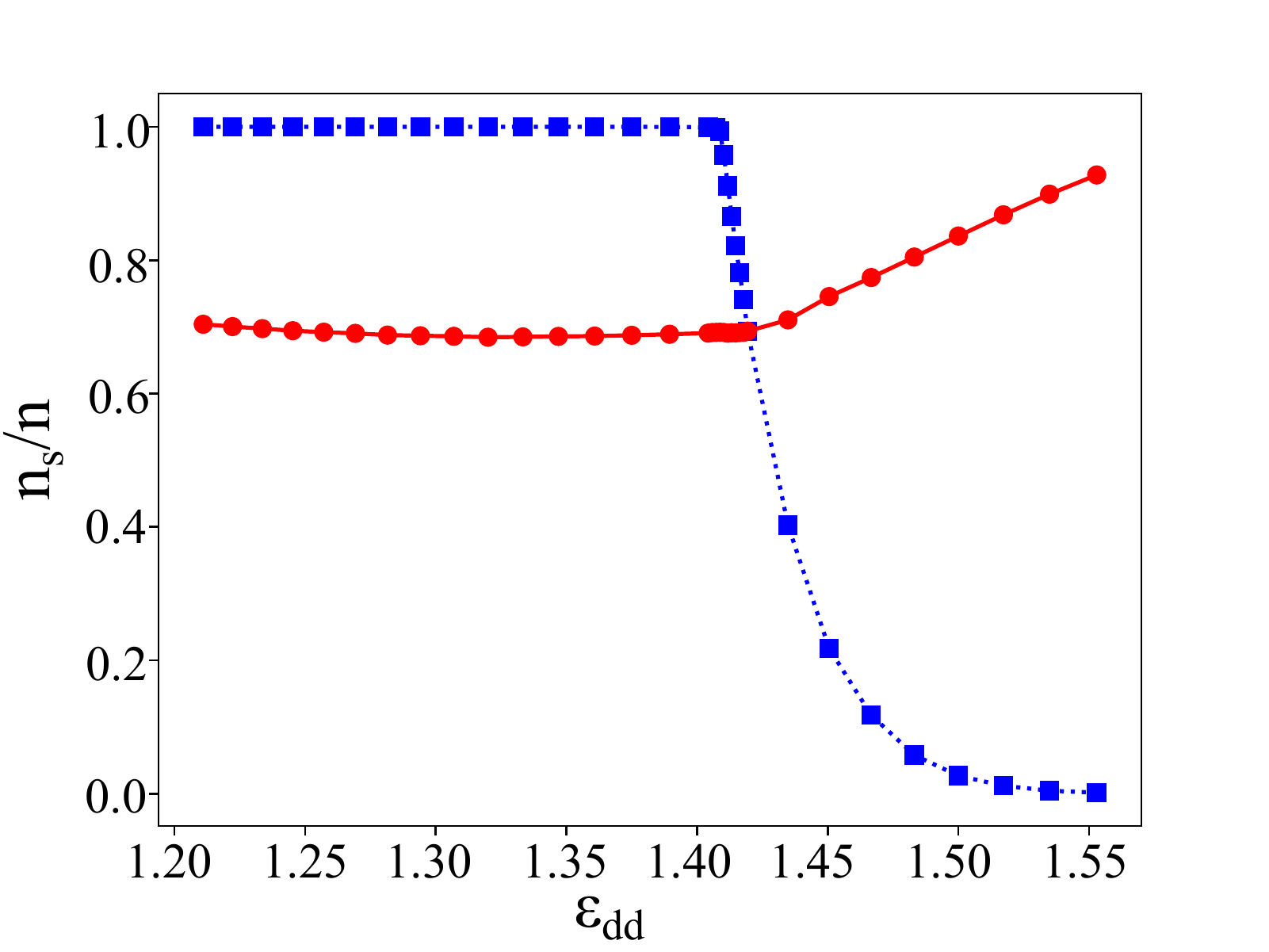}
    \caption{Estimate of the superfluid fraction of the edge region as function of $\epsilon_{dd}$, based on Leggett's variational formula \ref{leggett_supfrac} (blue squares), applied to the configuration described in figure \ref{circular_box}. Red circles represent the ratio between the number of atoms that settle on the edge and the total number of atoms in the system}
    \label{leggett}
\end{figure}
The estimate \ref{leggett_supfrac}, reported in Fig. \ref{leggett}  (blue squares), reveals a critical dependence on  $\epsilon_{dd}$, emphasizing the emergence of a phase transition between the superfluid and the supersolid phase at $\epsilon_{dd}=1.4$ and a transition  between the supersolid and the crystal phase, characterized by the vanishing of $n_S$, at $\epsilon_{dd}=1.55$. These values are very close to the critical values calculated for one dimensional tubular configurations imposing periodic boundary conditions \cite{Roccuzzo1}, after taking into account that in the edge  configuration discussed here the number of  atoms occupying  the ring increases with  $\epsilon_{dd}$, as shown in the same figure (red circles).
Such an increase is actually particularly important in the supersolid phase as a consequence of  the reduced value of the chemical potential, which favors the accumulation of dipoles on the density peaks, where the inter-atomic dipolar interaction is mainly attractive. 

The novel  configuration emerging in the box of circular shape discussed above is particularly attractive because in this case the boundary does not depend on the azimuthal coordinate and takes  the form of a  ring, where  the dipolar particles form a one-dimensional structure, well separated from the atoms in the bulk. This  provides  the interesting possibility of exploring superfluid and supersolid features in uniform one dimensional like configurations with periodic boundary conditions. 
\paragraph{Square box potential.} It is interesting to consider other forms of  boxes like, for example, the most familiar square box. This case was considered in \cite{Lu2010}  in  the deep superfluid phase and in the absence of beyond-mean-field effects. Here, we consider also  regimes where the mean field approach would yield instability and the LHY correction allows for the emergence of the supersolid and crystal phases.

\begin{figure}
    \includegraphics[width=0.5\textwidth]{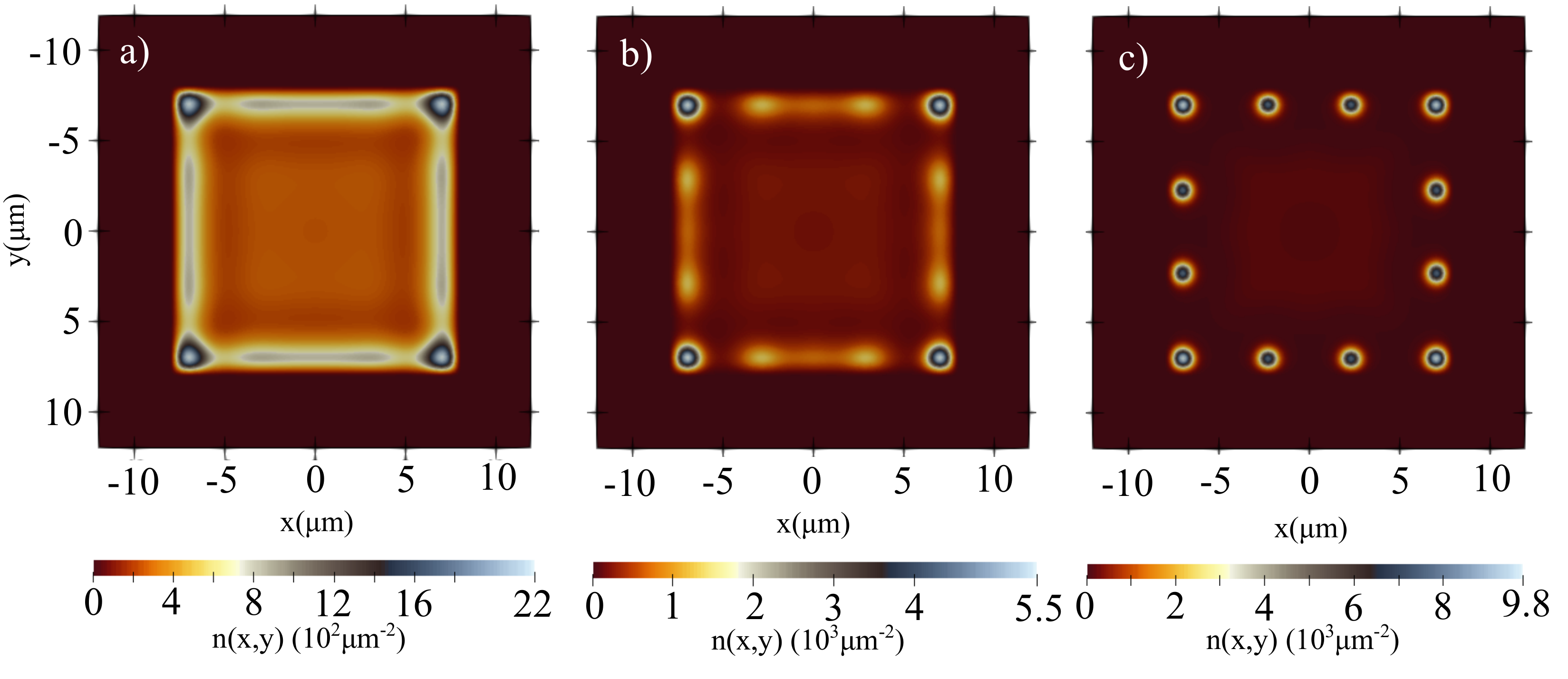}
    \caption{Ground state integrated density profiles $\mbox{n(x,y)=}\int\mbox{dz}|\Psi\mbox{(x,y,z)}|^2$ 
    for a gas of $10^5$ atoms of $^{164}$Dy confined in the polarization direction by a harmonic potential of frequency
    $\omega_z=(2\pi)100$Hz, and by a box potential in the x-y plain, with the shape of a square of side $L=16\mu m$. The value of $\epsilon_{dd}$ for panels a), b) and c) is, respectively, 1.32, 1.404 and 1.467. The
    height of the box is fixed to $\mbox{V}_0=100\hbar\omega_z$}
    \label{square_box}
\end{figure}

The results for the density profiles in the case of a 2D square box are reported in Fig. \ref{square_box} and  reveal  the same mechanism of accumulation of the density near the boundary already discussed in the case of a circular box. A major difference concerns the behavior of the density profile along the edge of the box. In fact the vertices of the square box become points of strong accumulation of dipoles, causing density modulations along the sides of the square, even for small values of $\epsilon_{dd}$ (see Fig. \ref{square_box} panel a), when the system is in the superfluid phase. The behavior of the density along each edge of the square configuration shares interesting analogies with the behavior exhibited by a quasi one-dimensional gas confined by a box potential (see Supplementary Materials).

The above discussion reveals that the presence of the square box makes the identification of the transition between the  superfluid, supersolid and crystal phases of the dipolar gas on the edge of the box more difficult than in the case of the circular box.


\paragraph{Bulk supersolidity.} The configurations discussed so far do not reveal the emergence of supersolid effects in the bulk region, because of the small value of the bulk density caused by the accumulation of dipoles near the boundary. In order to observe the bulk supersolidity one consequently needs to increase significantly the atom density, in such a way that the density in the central region remains large enough to ensure the appearance of a crystal quantum phase. In Fig. \ref{geometries}, we have considered  configurations  containing $N=2\times10^6$ atoms confined by a box potential in the transverse direction, with the shape of regular polygons (panels a-d) or circular (panel e), all with the same area (and hence the same number of atoms per unit surface). 
For the same value of $\epsilon_{dd}=1.36$, these configurations exhibit a supersolid structure in the bulk, characterized by the typical overlap between neighbouring density peaks, well separated from the edge region by a density dip. Despite the number of atoms and system size considered,  resulting in a large number of droplets ($\simeq 60)$, the symmetry of the supersolid lattice reflects the one of the confining potential, implying that surface effects hinder the possibility of reaching the thermodynamic limit, where the lattice is expected to be triangular or honeycomb \cite{Pohl2019}.
This can be qualitatively understood as a consequence of the long-range nature of the dipolar force and the formation of the edge. In fact, since the dipoles are in a mainly repulsive configuration, they tend to expand  towards the edge, where they acquire a density profile with the same shape of the confining potential; the droplets that form in the bulk also tend to repel each other, but their expansion is stopped by the repulsion of the edge, so that they are forced to arrange in lines parallel to the sides of the edge. This behavior is suppressed in an infinite system or in a harmonic trap, where the expansion of the gas is energetically unfavourable. 

It is worth noticing that the supersolid and crystal structures at the edge of the boundary, which are well visible in the configurations of Fig. \ref{circular_box} and \ref{square_box} (panels b and c), have disappeared in Fig. \ref{geometries} as a consequence of the high density acquired by the system near the boundary, caused by  the large value of $N$. As pointed out in \cite{Blakie_1d_1} and discussed also in \cite{hertkorn2021}, the density dependence of the critical value of the interaction parameter  $\epsilon_{dd}$, which  separates the superfluid from the supersolid phase, actually exhibits a characteristic non monotonic dependence (see also Fig. \ref{phase_diagram} in the Supplementary Materials). This implies that, for a properly fixed value of $\epsilon_{dd}$, if one increases  the density starting from small values, the system undergoes first  a phase transition from the superfluid to the supersolid (and eventually to the crystal) phase characterized by typical  density oscillations,  to come back again to the uniform superfluid phase at larger densities. Notice that this effect can also be observed with a smaller number of atoms, by confining the atoms in properly designed box potentials of smaller dimension. In fact, such density, although relatively high ($\simeq 10^{15}cm^{-3}$ for the edge configurations shown in Fig. \ref{geometries}), is still compatible with the usual stability conditions imposed by three-body recombination, suggesting the possibility of observing this effect in actual experiments.  


We have finally checked that the results presented in this work do not qualitatively change for different choices of
the parameters. In particular  we have considered  different values of the  transverse confinement
in the  interval $(2\pi)50\mbox{Hz}<\omega_y=\omega_z<(2\pi)150\mbox{Hz}$, and of system size and number of atoms. The actual choice of $\omega_{y,z}$ can however affect the value of the density in the central region, the critical value of $\epsilon_{dd}$ for the superfluid-supersolid phase transition, as well as the number of droplets which form in the supersolid phase, their relative distance being sensitive to the value of $\omega_z$ \cite{SantosRoton}. \\

 In conclusion, we have investigated  the ground state configurations of a dipolar Bose-Einstein condensed gas confined by a box potential. We have shown that the tendency of the density  to accumulate near the walls, as a consequence of the repulsion between aligned dipoles, favours the formation of novel quasi-1d configurations located at the edge of the box and  well separated from  the  atoms filling the bulk region. In the case of quasi-2d boxes of circular shape the edge configuration takes the characteristic  form of a ring, revealing clear  supersolid and crystal effects in a useful range of parameters. We have also shown that the geometry of the supersolid  in the bulk region reflects the shape of the confining potential even for very large systems, therefore  hindering the possibility of reaching the thermodynamic limit of  dipolar BEC's  using box potentials. Natural extension of this work concern the study of the non equilibrium behavior exhibited by dipolar gases in the novel ring configuration formed at the edge of the circular box. 

\paragraph*{Acknowledgement}
Useful discussions with G. Modugno and the members of the Firenze-Pisa dipolar group are
acknowledged. This  project has  received  funding from  Provincia Autonoma  di  Trento,  the  Q@TN
initiative  and  the  FIS$\hbar$ project of the Istituto Nazionale di Fisica Nucleare, and the
Italian MIUR under the PRIN2017 project CEnTraL.

\bibliography{biblio_meniscus.bib}

\begin{thebibliography}{54}%
\makeatletter
\providecommand \@ifxundefined [1]{%
 \@ifx{#1\undefined}
}%
\providecommand \@ifnum [1]{%
 \ifnum #1\expandafter \@firstoftwo
 \else \expandafter \@secondoftwo
 \fi
}%
\providecommand \@ifx [1]{%
 \ifx #1\expandafter \@firstoftwo
 \else \expandafter \@secondoftwo
 \fi
}%
\providecommand \natexlab [1]{#1}%
\providecommand \enquote  [1]{``#1''}%
\providecommand \bibnamefont  [1]{#1}%
\providecommand \bibfnamefont [1]{#1}%
\providecommand \citenamefont [1]{#1}%
\providecommand \href@noop [0]{\@secondoftwo}%
\providecommand \href [0]{\begingroup \@sanitize@url \@href}%
\providecommand \@href[1]{\@@startlink{#1}\@@href}%
\providecommand \@@href[1]{\endgroup#1\@@endlink}%
\providecommand \@sanitize@url [0]{\catcode `\\12\catcode `\$12\catcode
  `\&12\catcode `\#12\catcode `\^12\catcode `\_12\catcode `\%12\relax}%
\providecommand \@@startlink[1]{}%
\providecommand \@@endlink[0]{}%
\providecommand \url  [0]{\begingroup\@sanitize@url \@url }%
\providecommand \@url [1]{\endgroup\@href {#1}{\urlprefix }}%
\providecommand \urlprefix  [0]{URL }%
\providecommand \Eprint [0]{\href }%
\providecommand \doibase [0]{http://dx.doi.org/}%
\providecommand \selectlanguage [0]{\@gobble}%
\providecommand \bibinfo  [0]{\@secondoftwo}%
\providecommand \bibfield  [0]{\@secondoftwo}%
\providecommand \translation [1]{[#1]}%
\providecommand \BibitemOpen [0]{}%
\providecommand \bibitemStop [0]{}%
\providecommand \bibitemNoStop [0]{.\EOS\space}%
\providecommand \EOS [0]{\spacefactor3000\relax}%
\providecommand \BibitemShut  [1]{\csname bibitem#1\endcsname}%
\let\auto@bib@innerbib\@empty
\bibitem [{\citenamefont {Pitaevskii}\ and\ \citenamefont
  {Stringari}(2016)}]{BecBook2016}%
  \BibitemOpen
  \bibfield  {author} {\bibinfo {author} {\bibfnamefont {L.}~\bibnamefont
  {Pitaevskii}}\ and\ \bibinfo {author} {\bibfnamefont {S.}~\bibnamefont
  {Stringari}},\ }\href@noop {} {\emph {\bibinfo {title} {Bose-Einstein
  condensation and superfluidity}}}\ (\bibinfo  {publisher} {Oxford University
  Press},\ \bibinfo {year} {2016})\BibitemShut {NoStop}%
\bibitem [{\citenamefont {Stringari}(1996)}]{stringari96}%
  \BibitemOpen
  \bibfield  {author} {\bibinfo {author} {\bibfnamefont {S.}~\bibnamefont
  {Stringari}},\ }\href {\doibase 10.1103/PhysRevLett.77.2360} {\bibfield
  {journal} {\bibinfo  {journal} {Phys. Rev. Lett.}\ }\textbf {\bibinfo
  {volume} {77}},\ \bibinfo {pages} {2360} (\bibinfo {year}
  {1996})}\BibitemShut {NoStop}%
\bibitem [{\citenamefont {Mewes}\ \emph {et~al.}(1996)\citenamefont {Mewes},
  \citenamefont {Andrews}, \citenamefont {van Druten}, \citenamefont {Kurn},
  \citenamefont {Durfee}, \citenamefont {Townsend},\ and\ \citenamefont
  {Ketterle}}]{mewes96}%
  \BibitemOpen
  \bibfield  {author} {\bibinfo {author} {\bibfnamefont {M.-O.}\ \bibnamefont
  {Mewes}}, \bibinfo {author} {\bibfnamefont {M.~R.}\ \bibnamefont {Andrews}},
  \bibinfo {author} {\bibfnamefont {N.~J.}\ \bibnamefont {van Druten}},
  \bibinfo {author} {\bibfnamefont {D.~M.}\ \bibnamefont {Kurn}}, \bibinfo
  {author} {\bibfnamefont {D.~S.}\ \bibnamefont {Durfee}}, \bibinfo {author}
  {\bibfnamefont {C.~G.}\ \bibnamefont {Townsend}}, \ and\ \bibinfo {author}
  {\bibfnamefont {W.}~\bibnamefont {Ketterle}},\ }\href {\doibase
  10.1103/PhysRevLett.77.988} {\bibfield  {journal} {\bibinfo  {journal} {Phys.
  Rev. Lett.}\ }\textbf {\bibinfo {volume} {77}},\ \bibinfo {pages} {988}
  (\bibinfo {year} {1996})}\BibitemShut {NoStop}%
\bibitem [{\citenamefont {Gu\'ery-Odelin}\ and\ \citenamefont
  {Stringari}(1999)}]{guery_odelin1999}%
  \BibitemOpen
  \bibfield  {author} {\bibinfo {author} {\bibfnamefont {D.}~\bibnamefont
  {Gu\'ery-Odelin}}\ and\ \bibinfo {author} {\bibfnamefont {S.}~\bibnamefont
  {Stringari}},\ }\href {\doibase 10.1103/PhysRevLett.83.4452} {\bibfield
  {journal} {\bibinfo  {journal} {Phys. Rev. Lett.}\ }\textbf {\bibinfo
  {volume} {83}},\ \bibinfo {pages} {4452} (\bibinfo {year}
  {1999})}\BibitemShut {NoStop}%
\bibitem [{\citenamefont {Marag\`o}\ \emph {et~al.}(2000)\citenamefont
  {Marag\`o}, \citenamefont {Hopkins}, \citenamefont {Arlt}, \citenamefont
  {Hodby}, \citenamefont {Hechenblaikner},\ and\ \citenamefont
  {Foot}}]{marago2000}%
  \BibitemOpen
  \bibfield  {author} {\bibinfo {author} {\bibfnamefont {O.~M.}\ \bibnamefont
  {Marag\`o}}, \bibinfo {author} {\bibfnamefont {S.~A.}\ \bibnamefont
  {Hopkins}}, \bibinfo {author} {\bibfnamefont {J.}~\bibnamefont {Arlt}},
  \bibinfo {author} {\bibfnamefont {E.}~\bibnamefont {Hodby}}, \bibinfo
  {author} {\bibfnamefont {G.}~\bibnamefont {Hechenblaikner}}, \ and\ \bibinfo
  {author} {\bibfnamefont {C.~J.}\ \bibnamefont {Foot}},\ }\href {\doibase
  10.1103/PhysRevLett.84.2056} {\bibfield  {journal} {\bibinfo  {journal}
  {Phys. Rev. Lett.}\ }\textbf {\bibinfo {volume} {84}},\ \bibinfo {pages}
  {2056} (\bibinfo {year} {2000})}\BibitemShut {NoStop}%
\bibitem [{\citenamefont {Rossi}\ \emph {et~al.}(2016)\citenamefont {Rossi},
  \citenamefont {Dubessy}, \citenamefont {Merloti}, \citenamefont {de~Goër~de
  Herve}, \citenamefont {Badr}, \citenamefont {Perrin}, \citenamefont
  {Longchambon},\ and\ \citenamefont {Perrin}}]{Rossi2016}%
  \BibitemOpen
  \bibfield  {author} {\bibinfo {author} {\bibfnamefont {C.~D.}\ \bibnamefont
  {Rossi}}, \bibinfo {author} {\bibfnamefont {R.}~\bibnamefont {Dubessy}},
  \bibinfo {author} {\bibfnamefont {K.}~\bibnamefont {Merloti}}, \bibinfo
  {author} {\bibfnamefont {M.}~\bibnamefont {de~Goër~de Herve}}, \bibinfo
  {author} {\bibfnamefont {T.}~\bibnamefont {Badr}}, \bibinfo {author}
  {\bibfnamefont {A.}~\bibnamefont {Perrin}}, \bibinfo {author} {\bibfnamefont
  {L.}~\bibnamefont {Longchambon}}, \ and\ \bibinfo {author} {\bibfnamefont
  {H.}~\bibnamefont {Perrin}},\ }\href {\doibase 10.1088/1367-2630/18/6/062001}
  {\bibfield  {journal} {\bibinfo  {journal} {New Journal of Physics}\ }\textbf
  {\bibinfo {volume} {18}},\ \bibinfo {pages} {062001} (\bibinfo {year}
  {2016})}\BibitemShut {NoStop}%
\bibitem [{\citenamefont {Madison}\ \emph {et~al.}(2000)\citenamefont
  {Madison}, \citenamefont {Chevy}, \citenamefont {Wohlleben},\ and\
  \citenamefont {Dalibard}}]{madison2000}%
  \BibitemOpen
  \bibfield  {author} {\bibinfo {author} {\bibfnamefont {K.~W.}\ \bibnamefont
  {Madison}}, \bibinfo {author} {\bibfnamefont {F.}~\bibnamefont {Chevy}},
  \bibinfo {author} {\bibfnamefont {W.}~\bibnamefont {Wohlleben}}, \ and\
  \bibinfo {author} {\bibfnamefont {J.}~\bibnamefont {Dalibard}},\ }\href
  {\doibase 10.1103/PhysRevLett.84.806} {\bibfield  {journal} {\bibinfo
  {journal} {Phys. Rev. Lett.}\ }\textbf {\bibinfo {volume} {84}},\ \bibinfo
  {pages} {806} (\bibinfo {year} {2000})}\BibitemShut {NoStop}%
\bibitem [{\citenamefont {Abo-Shaeer}\ \emph {et~al.}(2001)\citenamefont
  {Abo-Shaeer}, \citenamefont {Raman}, \citenamefont {Vogels},\ and\
  \citenamefont {Ketterle}}]{AboShaeer2001}%
  \BibitemOpen
  \bibfield  {author} {\bibinfo {author} {\bibfnamefont {J.~R.}\ \bibnamefont
  {Abo-Shaeer}}, \bibinfo {author} {\bibfnamefont {C.}~\bibnamefont {Raman}},
  \bibinfo {author} {\bibfnamefont {J.~M.}\ \bibnamefont {Vogels}}, \ and\
  \bibinfo {author} {\bibfnamefont {W.}~\bibnamefont {Ketterle}},\ }\href
  {\doibase 10.1126/science.1060182} {\bibfield  {journal} {\bibinfo  {journal}
  {Science}\ }\textbf {\bibinfo {volume} {292}},\ \bibinfo {pages} {476}
  (\bibinfo {year} {2001})},\ \Eprint
  {http://arxiv.org/abs/https://science.sciencemag.org/content/292/5516/476.full.pdf}
  {https://science.sciencemag.org/content/292/5516/476.full.pdf} \BibitemShut
  {NoStop}%
\bibitem [{\citenamefont {Haljan}\ \emph {et~al.}(2001)\citenamefont {Haljan},
  \citenamefont {Coddington}, \citenamefont {Engels},\ and\ \citenamefont
  {Cornell}}]{Haljan2001}%
  \BibitemOpen
  \bibfield  {author} {\bibinfo {author} {\bibfnamefont {P.~C.}\ \bibnamefont
  {Haljan}}, \bibinfo {author} {\bibfnamefont {I.}~\bibnamefont {Coddington}},
  \bibinfo {author} {\bibfnamefont {P.}~\bibnamefont {Engels}}, \ and\ \bibinfo
  {author} {\bibfnamefont {E.~A.}\ \bibnamefont {Cornell}},\ }\href {\doibase
  10.1103/PhysRevLett.87.210403} {\bibfield  {journal} {\bibinfo  {journal}
  {Phys. Rev. Lett.}\ }\textbf {\bibinfo {volume} {87}},\ \bibinfo {pages}
  {210403} (\bibinfo {year} {2001})}\BibitemShut {NoStop}%
\bibitem [{\citenamefont {Meyrath}\ \emph {et~al.}(2005)\citenamefont
  {Meyrath}, \citenamefont {Schreck}, \citenamefont {Hanssen}, \citenamefont
  {Chuu},\ and\ \citenamefont {Raizen}}]{Raizen2005}%
  \BibitemOpen
  \bibfield  {author} {\bibinfo {author} {\bibfnamefont {T.~P.}\ \bibnamefont
  {Meyrath}}, \bibinfo {author} {\bibfnamefont {F.}~\bibnamefont {Schreck}},
  \bibinfo {author} {\bibfnamefont {J.~L.}\ \bibnamefont {Hanssen}}, \bibinfo
  {author} {\bibfnamefont {C.-S.}\ \bibnamefont {Chuu}}, \ and\ \bibinfo
  {author} {\bibfnamefont {M.~G.}\ \bibnamefont {Raizen}},\ }\href {\doibase
  10.1103/PhysRevA.71.041604} {\bibfield  {journal} {\bibinfo  {journal} {Phys.
  Rev. A}\ }\textbf {\bibinfo {volume} {71}},\ \bibinfo {pages} {041604}
  (\bibinfo {year} {2005})}\BibitemShut {NoStop}%
\bibitem [{\citenamefont {Gaunt}\ \emph {et~al.}(2013)\citenamefont {Gaunt},
  \citenamefont {Schmidutz}, \citenamefont {Gotlibovych}, \citenamefont
  {Smith},\ and\ \citenamefont {Hadzibabic}}]{Hadzibabic2013}%
  \BibitemOpen
  \bibfield  {author} {\bibinfo {author} {\bibfnamefont {A.~L.}\ \bibnamefont
  {Gaunt}}, \bibinfo {author} {\bibfnamefont {T.~F.}\ \bibnamefont
  {Schmidutz}}, \bibinfo {author} {\bibfnamefont {I.}~\bibnamefont
  {Gotlibovych}}, \bibinfo {author} {\bibfnamefont {R.~P.}\ \bibnamefont
  {Smith}}, \ and\ \bibinfo {author} {\bibfnamefont {Z.}~\bibnamefont
  {Hadzibabic}},\ }\href {\doibase 10.1103/PhysRevLett.110.200406} {\bibfield
  {journal} {\bibinfo  {journal} {Phys. Rev. Lett.}\ }\textbf {\bibinfo
  {volume} {110}},\ \bibinfo {pages} {200406} (\bibinfo {year}
  {2013})}\BibitemShut {NoStop}%
\bibitem [{\citenamefont {Gupta}\ \emph {et~al.}(2005)\citenamefont {Gupta},
  \citenamefont {Murch}, \citenamefont {Moore}, \citenamefont {Purdy},\ and\
  \citenamefont {Stamper-Kurn}}]{Gupta2005}%
  \BibitemOpen
  \bibfield  {author} {\bibinfo {author} {\bibfnamefont {S.}~\bibnamefont
  {Gupta}}, \bibinfo {author} {\bibfnamefont {K.~W.}\ \bibnamefont {Murch}},
  \bibinfo {author} {\bibfnamefont {K.~L.}\ \bibnamefont {Moore}}, \bibinfo
  {author} {\bibfnamefont {T.~P.}\ \bibnamefont {Purdy}}, \ and\ \bibinfo
  {author} {\bibfnamefont {D.~M.}\ \bibnamefont {Stamper-Kurn}},\ }\href
  {\doibase 10.1103/PhysRevLett.95.143201} {\bibfield  {journal} {\bibinfo
  {journal} {Phys. Rev. Lett.}\ }\textbf {\bibinfo {volume} {95}},\ \bibinfo
  {pages} {143201} (\bibinfo {year} {2005})}\BibitemShut {NoStop}%
\bibitem [{\citenamefont {Navon}\ \emph {et~al.}(2015)\citenamefont {Navon},
  \citenamefont {Gaunt}, \citenamefont {Smith},\ and\ \citenamefont
  {Hadzibabic}}]{Navon167}%
  \BibitemOpen
  \bibfield  {author} {\bibinfo {author} {\bibfnamefont {N.}~\bibnamefont
  {Navon}}, \bibinfo {author} {\bibfnamefont {A.~L.}\ \bibnamefont {Gaunt}},
  \bibinfo {author} {\bibfnamefont {R.~P.}\ \bibnamefont {Smith}}, \ and\
  \bibinfo {author} {\bibfnamefont {Z.}~\bibnamefont {Hadzibabic}},\ }\href
  {\doibase 10.1126/science.1258676} {\bibfield  {journal} {\bibinfo  {journal}
  {Science}\ }\textbf {\bibinfo {volume} {347}},\ \bibinfo {pages} {167}
  (\bibinfo {year} {2015})},\ \Eprint
  {http://arxiv.org/abs/https://science.sciencemag.org/content/347/6218/167.full.pdf}
  {https://science.sciencemag.org/content/347/6218/167.full.pdf} \BibitemShut
  {NoStop}%
\bibitem [{\citenamefont {Chomaz}\ \emph {et~al.}(2015)\citenamefont {Chomaz},
  \citenamefont {Corman}, \citenamefont {Bienaim{\'e}}, \citenamefont
  {Desbuquois}, \citenamefont {Weitenberg}, \citenamefont {Nascimb{\`e}ne},
  \citenamefont {Beugnon},\ and\ \citenamefont {Dalibard}}]{Chomaz2015}%
  \BibitemOpen
  \bibfield  {author} {\bibinfo {author} {\bibfnamefont {L.}~\bibnamefont
  {Chomaz}}, \bibinfo {author} {\bibfnamefont {L.}~\bibnamefont {Corman}},
  \bibinfo {author} {\bibfnamefont {T.}~\bibnamefont {Bienaim{\'e}}}, \bibinfo
  {author} {\bibfnamefont {R.}~\bibnamefont {Desbuquois}}, \bibinfo {author}
  {\bibfnamefont {C.}~\bibnamefont {Weitenberg}}, \bibinfo {author}
  {\bibfnamefont {S.}~\bibnamefont {Nascimb{\`e}ne}}, \bibinfo {author}
  {\bibfnamefont {J.}~\bibnamefont {Beugnon}}, \ and\ \bibinfo {author}
  {\bibfnamefont {J.}~\bibnamefont {Dalibard}},\ }\href {\doibase
  10.1038/ncomms7162} {\bibfield  {journal} {\bibinfo  {journal} {Nature
  Communications}\ }\textbf {\bibinfo {volume} {6}},\ \bibinfo {pages} {6162}
  (\bibinfo {year} {2015})}\BibitemShut {NoStop}%
\bibitem [{\citenamefont {Navon}\ \emph {et~al.}(2016)\citenamefont {Navon},
  \citenamefont {Gaunt}, \citenamefont {Smith},\ and\ \citenamefont
  {Hadzibabic}}]{Navon2016}%
  \BibitemOpen
  \bibfield  {author} {\bibinfo {author} {\bibfnamefont {N.}~\bibnamefont
  {Navon}}, \bibinfo {author} {\bibfnamefont {A.~L.}\ \bibnamefont {Gaunt}},
  \bibinfo {author} {\bibfnamefont {R.~P.}\ \bibnamefont {Smith}}, \ and\
  \bibinfo {author} {\bibfnamefont {Z.}~\bibnamefont {Hadzibabic}},\ }\href
  {\doibase 10.1038/nature20114} {\bibfield  {journal} {\bibinfo  {journal}
  {Nature}\ }\textbf {\bibinfo {volume} {539}},\ \bibinfo {pages} {72}
  (\bibinfo {year} {2016})}\BibitemShut {NoStop}%
\bibitem [{\citenamefont {Ville}\ \emph {et~al.}(2017)\citenamefont {Ville},
  \citenamefont {Bienaim\'e}, \citenamefont {Saint-Jalm}, \citenamefont
  {Corman}, \citenamefont {Aidelsburger}, \citenamefont {Chomaz}, \citenamefont
  {Kleinlein}, \citenamefont {Perconte}, \citenamefont {Nascimb\`ene},
  \citenamefont {Dalibard},\ and\ \citenamefont {Beugnon}}]{Ville2017}%
  \BibitemOpen
  \bibfield  {author} {\bibinfo {author} {\bibfnamefont {J.~L.}\ \bibnamefont
  {Ville}}, \bibinfo {author} {\bibfnamefont {T.}~\bibnamefont {Bienaim\'e}},
  \bibinfo {author} {\bibfnamefont {R.}~\bibnamefont {Saint-Jalm}}, \bibinfo
  {author} {\bibfnamefont {L.}~\bibnamefont {Corman}}, \bibinfo {author}
  {\bibfnamefont {M.}~\bibnamefont {Aidelsburger}}, \bibinfo {author}
  {\bibfnamefont {L.}~\bibnamefont {Chomaz}}, \bibinfo {author} {\bibfnamefont
  {K.}~\bibnamefont {Kleinlein}}, \bibinfo {author} {\bibfnamefont
  {D.}~\bibnamefont {Perconte}}, \bibinfo {author} {\bibfnamefont
  {S.}~\bibnamefont {Nascimb\`ene}}, \bibinfo {author} {\bibfnamefont
  {J.}~\bibnamefont {Dalibard}}, \ and\ \bibinfo {author} {\bibfnamefont
  {J.}~\bibnamefont {Beugnon}},\ }\href {\doibase 10.1103/PhysRevA.95.013632}
  {\bibfield  {journal} {\bibinfo  {journal} {Phys. Rev. A}\ }\textbf {\bibinfo
  {volume} {95}},\ \bibinfo {pages} {013632} (\bibinfo {year}
  {2017})}\BibitemShut {NoStop}%
\bibitem [{\citenamefont {Ville}\ \emph {et~al.}(2018)\citenamefont {Ville},
  \citenamefont {Saint-Jalm}, \citenamefont {Le~Cerf}, \citenamefont
  {Aidelsburger}, \citenamefont {Nascimb\`ene}, \citenamefont {Dalibard},\ and\
  \citenamefont {Beugnon}}]{Ville2018}%
  \BibitemOpen
  \bibfield  {author} {\bibinfo {author} {\bibfnamefont {J.~L.}\ \bibnamefont
  {Ville}}, \bibinfo {author} {\bibfnamefont {R.}~\bibnamefont {Saint-Jalm}},
  \bibinfo {author} {\bibfnamefont {E.}~\bibnamefont {Le~Cerf}}, \bibinfo
  {author} {\bibfnamefont {M.}~\bibnamefont {Aidelsburger}}, \bibinfo {author}
  {\bibfnamefont {S.}~\bibnamefont {Nascimb\`ene}}, \bibinfo {author}
  {\bibfnamefont {J.}~\bibnamefont {Dalibard}}, \ and\ \bibinfo {author}
  {\bibfnamefont {J.}~\bibnamefont {Beugnon}},\ }\href {\doibase
  10.1103/PhysRevLett.121.145301} {\bibfield  {journal} {\bibinfo  {journal}
  {Phys. Rev. Lett.}\ }\textbf {\bibinfo {volume} {121}},\ \bibinfo {pages}
  {145301} (\bibinfo {year} {2018})}\BibitemShut {NoStop}%
\bibitem [{\citenamefont {Griesmaier}\ \emph {et~al.}(2005)\citenamefont
  {Griesmaier}, \citenamefont {Werner}, \citenamefont {Hensler}, \citenamefont
  {Stuhler},\ and\ \citenamefont {Pfau}}]{BEC_Cr}%
  \BibitemOpen
  \bibfield  {author} {\bibinfo {author} {\bibfnamefont {A.}~\bibnamefont
  {Griesmaier}}, \bibinfo {author} {\bibfnamefont {J.}~\bibnamefont {Werner}},
  \bibinfo {author} {\bibfnamefont {S.}~\bibnamefont {Hensler}}, \bibinfo
  {author} {\bibfnamefont {J.}~\bibnamefont {Stuhler}}, \ and\ \bibinfo
  {author} {\bibfnamefont {T.}~\bibnamefont {Pfau}},\ }\href {\doibase
  10.1103/PhysRevLett.94.160401} {\bibfield  {journal} {\bibinfo  {journal}
  {Phys. Rev. Lett.}\ }\textbf {\bibinfo {volume} {94}},\ \bibinfo {pages}
  {160401} (\bibinfo {year} {2005})}\BibitemShut {NoStop}%
\bibitem [{\citenamefont {Lucioni}\ \emph {et~al.}(2018)\citenamefont
  {Lucioni}, \citenamefont {Tanzi}, \citenamefont {Fregosi}, \citenamefont
  {Catani}, \citenamefont {Gozzini}, \citenamefont {Inguscio}, \citenamefont
  {Fioretti}, \citenamefont {Gabbanini},\ and\ \citenamefont
  {Modugno}}]{BEC_Dy162}%
  \BibitemOpen
  \bibfield  {author} {\bibinfo {author} {\bibfnamefont {E.}~\bibnamefont
  {Lucioni}}, \bibinfo {author} {\bibfnamefont {L.}~\bibnamefont {Tanzi}},
  \bibinfo {author} {\bibfnamefont {A.}~\bibnamefont {Fregosi}}, \bibinfo
  {author} {\bibfnamefont {J.}~\bibnamefont {Catani}}, \bibinfo {author}
  {\bibfnamefont {S.}~\bibnamefont {Gozzini}}, \bibinfo {author} {\bibfnamefont
  {M.}~\bibnamefont {Inguscio}}, \bibinfo {author} {\bibfnamefont
  {A.}~\bibnamefont {Fioretti}}, \bibinfo {author} {\bibfnamefont
  {C.}~\bibnamefont {Gabbanini}}, \ and\ \bibinfo {author} {\bibfnamefont
  {G.}~\bibnamefont {Modugno}},\ }\href {\doibase 10.1103/PhysRevA.97.060701}
  {\bibfield  {journal} {\bibinfo  {journal} {Phys. Rev. A}\ }\textbf {\bibinfo
  {volume} {97}},\ \bibinfo {pages} {060701} (\bibinfo {year}
  {2018})}\BibitemShut {NoStop}%
\bibitem [{\citenamefont {Lu}\ \emph {et~al.}(2011)\citenamefont {Lu},
  \citenamefont {Burdick}, \citenamefont {Youn},\ and\ \citenamefont
  {Lev}}]{BEC_Dy164}%
  \BibitemOpen
  \bibfield  {author} {\bibinfo {author} {\bibfnamefont {M.}~\bibnamefont
  {Lu}}, \bibinfo {author} {\bibfnamefont {N.~Q.}\ \bibnamefont {Burdick}},
  \bibinfo {author} {\bibfnamefont {S.~H.}\ \bibnamefont {Youn}}, \ and\
  \bibinfo {author} {\bibfnamefont {B.~L.}\ \bibnamefont {Lev}},\ }\href
  {\doibase 10.1103/PhysRevLett.107.190401} {\bibfield  {journal} {\bibinfo
  {journal} {Phys. Rev. Lett.}\ }\textbf {\bibinfo {volume} {107}},\ \bibinfo
  {pages} {190401} (\bibinfo {year} {2011})}\BibitemShut {NoStop}%
\bibitem [{\citenamefont {Aikawa}\ \emph {et~al.}(2012)\citenamefont {Aikawa},
  \citenamefont {Frisch}, \citenamefont {Mark}, \citenamefont {Baier},
  \citenamefont {Rietzler}, \citenamefont {Grimm},\ and\ \citenamefont
  {Ferlaino}}]{BEC_Er}%
  \BibitemOpen
  \bibfield  {author} {\bibinfo {author} {\bibfnamefont {K.}~\bibnamefont
  {Aikawa}}, \bibinfo {author} {\bibfnamefont {A.}~\bibnamefont {Frisch}},
  \bibinfo {author} {\bibfnamefont {M.}~\bibnamefont {Mark}}, \bibinfo {author}
  {\bibfnamefont {S.}~\bibnamefont {Baier}}, \bibinfo {author} {\bibfnamefont
  {A.}~\bibnamefont {Rietzler}}, \bibinfo {author} {\bibfnamefont
  {R.}~\bibnamefont {Grimm}}, \ and\ \bibinfo {author} {\bibfnamefont
  {F.}~\bibnamefont {Ferlaino}},\ }\href {\doibase
  10.1103/PhysRevLett.108.210401} {\bibfield  {journal} {\bibinfo  {journal}
  {Phys. Rev. Lett.}\ }\textbf {\bibinfo {volume} {108}},\ \bibinfo {pages}
  {210401} (\bibinfo {year} {2012})}\BibitemShut {NoStop}%
\bibitem [{\citenamefont {Koch}\ \emph {et~al.}(2008)\citenamefont {Koch},
  \citenamefont {Lahaye}, \citenamefont {Metz}, \citenamefont {Fr{\"o}hlich},
  \citenamefont {Griesmaier},\ and\ \citenamefont {Pfau}}]{Koch2008}%
  \BibitemOpen
  \bibfield  {author} {\bibinfo {author} {\bibfnamefont {T.}~\bibnamefont
  {Koch}}, \bibinfo {author} {\bibfnamefont {T.}~\bibnamefont {Lahaye}},
  \bibinfo {author} {\bibfnamefont {J.}~\bibnamefont {Metz}}, \bibinfo {author}
  {\bibfnamefont {B.}~\bibnamefont {Fr{\"o}hlich}}, \bibinfo {author}
  {\bibfnamefont {A.}~\bibnamefont {Griesmaier}}, \ and\ \bibinfo {author}
  {\bibfnamefont {T.}~\bibnamefont {Pfau}},\ }\href {\doibase 10.1038/nphys887}
  {\bibfield  {journal} {\bibinfo  {journal} {Nature Physics}\ }\textbf
  {\bibinfo {volume} {4}},\ \bibinfo {pages} {218} (\bibinfo {year}
  {2008})}\BibitemShut {NoStop}%
\bibitem [{\citenamefont {Santos}\ \emph {et~al.}(2003)\citenamefont {Santos},
  \citenamefont {Shlyapnikov},\ and\ \citenamefont {Lewenstein}}]{SantosRoton}%
  \BibitemOpen
  \bibfield  {author} {\bibinfo {author} {\bibfnamefont {L.}~\bibnamefont
  {Santos}}, \bibinfo {author} {\bibfnamefont {G.~V.}\ \bibnamefont
  {Shlyapnikov}}, \ and\ \bibinfo {author} {\bibfnamefont {M.}~\bibnamefont
  {Lewenstein}},\ }\href {\doibase 10.1103/PhysRevLett.90.250403} {\bibfield
  {journal} {\bibinfo  {journal} {Phys. Rev. Lett.}\ }\textbf {\bibinfo
  {volume} {90}},\ \bibinfo {pages} {250403} (\bibinfo {year}
  {2003})}\BibitemShut {NoStop}%
\bibitem [{\citenamefont {Roccuzzo}\ and\ \citenamefont
  {Ancilotto}(2019)}]{Roccuzzo1}%
  \BibitemOpen
  \bibfield  {author} {\bibinfo {author} {\bibfnamefont {S.~M.}\ \bibnamefont
  {Roccuzzo}}\ and\ \bibinfo {author} {\bibfnamefont {F.}~\bibnamefont
  {Ancilotto}},\ }\href {\doibase 10.1103/PhysRevA.99.041601} {\bibfield
  {journal} {\bibinfo  {journal} {Phys. Rev. A}\ }\textbf {\bibinfo {volume}
  {99}},\ \bibinfo {pages} {041601} (\bibinfo {year} {2019})}\BibitemShut
  {NoStop}%
\bibitem [{\citenamefont {Chomaz}\ \emph {et~al.}(2018)\citenamefont {Chomaz},
  \citenamefont {van Bijnen}, \citenamefont {Petter}, \citenamefont {Faraoni},
  \citenamefont {Baier}, \citenamefont {Becher}, \citenamefont {Mark},
  \citenamefont {W{\"a}chtler}, \citenamefont {Santos},\ and\ \citenamefont
  {Ferlaino}}]{Chomaz2018}%
  \BibitemOpen
  \bibfield  {author} {\bibinfo {author} {\bibfnamefont {L.}~\bibnamefont
  {Chomaz}}, \bibinfo {author} {\bibfnamefont {R.~M.~W.}\ \bibnamefont {van
  Bijnen}}, \bibinfo {author} {\bibfnamefont {D.}~\bibnamefont {Petter}},
  \bibinfo {author} {\bibfnamefont {G.}~\bibnamefont {Faraoni}}, \bibinfo
  {author} {\bibfnamefont {S.}~\bibnamefont {Baier}}, \bibinfo {author}
  {\bibfnamefont {J.~H.}\ \bibnamefont {Becher}}, \bibinfo {author}
  {\bibfnamefont {M.~J.}\ \bibnamefont {Mark}}, \bibinfo {author}
  {\bibfnamefont {F.}~\bibnamefont {W{\"a}chtler}}, \bibinfo {author}
  {\bibfnamefont {L.}~\bibnamefont {Santos}}, \ and\ \bibinfo {author}
  {\bibfnamefont {F.}~\bibnamefont {Ferlaino}},\ }\href {\doibase
  10.1038/s41567-018-0054-7} {\bibfield  {journal} {\bibinfo  {journal} {Nature
  Physics}\ }\textbf {\bibinfo {volume} {14}},\ \bibinfo {pages} {442}
  (\bibinfo {year} {2018})}\BibitemShut {NoStop}%
\bibitem [{\citenamefont {Petter}\ \emph {et~al.}(2019)\citenamefont {Petter},
  \citenamefont {Natale}, \citenamefont {van Bijnen}, \citenamefont
  {Patscheider}, \citenamefont {Mark}, \citenamefont {Chomaz},\ and\
  \citenamefont {Ferlaino}}]{Petter2019}%
  \BibitemOpen
  \bibfield  {author} {\bibinfo {author} {\bibfnamefont {D.}~\bibnamefont
  {Petter}}, \bibinfo {author} {\bibfnamefont {G.}~\bibnamefont {Natale}},
  \bibinfo {author} {\bibfnamefont {R.~M.~W.}\ \bibnamefont {van Bijnen}},
  \bibinfo {author} {\bibfnamefont {A.}~\bibnamefont {Patscheider}}, \bibinfo
  {author} {\bibfnamefont {M.~J.}\ \bibnamefont {Mark}}, \bibinfo {author}
  {\bibfnamefont {L.}~\bibnamefont {Chomaz}}, \ and\ \bibinfo {author}
  {\bibfnamefont {F.}~\bibnamefont {Ferlaino}},\ }\href {\doibase
  10.1103/PhysRevLett.122.183401} {\bibfield  {journal} {\bibinfo  {journal}
  {Phys. Rev. Lett.}\ }\textbf {\bibinfo {volume} {122}},\ \bibinfo {pages}
  {183401} (\bibinfo {year} {2019})}\BibitemShut {NoStop}%
\bibitem [{\citenamefont {Ferrier-Barbut}\ \emph
  {et~al.}(2016{\natexlab{a}})\citenamefont {Ferrier-Barbut}, \citenamefont
  {Kadau}, \citenamefont {Schmitt}, \citenamefont {Wenzel},\ and\ \citenamefont
  {Pfau}}]{ferrier_barbut2016}%
  \BibitemOpen
  \bibfield  {author} {\bibinfo {author} {\bibfnamefont {I.}~\bibnamefont
  {Ferrier-Barbut}}, \bibinfo {author} {\bibfnamefont {H.}~\bibnamefont
  {Kadau}}, \bibinfo {author} {\bibfnamefont {M.}~\bibnamefont {Schmitt}},
  \bibinfo {author} {\bibfnamefont {M.}~\bibnamefont {Wenzel}}, \ and\ \bibinfo
  {author} {\bibfnamefont {T.}~\bibnamefont {Pfau}},\ }\href {\doibase
  10.1103/PhysRevLett.116.215301} {\bibfield  {journal} {\bibinfo  {journal}
  {Phys. Rev. Lett.}\ }\textbf {\bibinfo {volume} {116}},\ \bibinfo {pages}
  {215301} (\bibinfo {year} {2016}{\natexlab{a}})}\BibitemShut {NoStop}%
\bibitem [{\citenamefont {Kadau}\ \emph {et~al.}(2016)\citenamefont {Kadau},
  \citenamefont {Schmitt}, \citenamefont {Wenzel}, \citenamefont {Wink},
  \citenamefont {Maier}, \citenamefont {Ferrier-Barbut},\ and\ \citenamefont
  {Pfau}}]{Kadau2016}%
  \BibitemOpen
  \bibfield  {author} {\bibinfo {author} {\bibfnamefont {H.}~\bibnamefont
  {Kadau}}, \bibinfo {author} {\bibfnamefont {M.}~\bibnamefont {Schmitt}},
  \bibinfo {author} {\bibfnamefont {M.}~\bibnamefont {Wenzel}}, \bibinfo
  {author} {\bibfnamefont {C.}~\bibnamefont {Wink}}, \bibinfo {author}
  {\bibfnamefont {T.}~\bibnamefont {Maier}}, \bibinfo {author} {\bibfnamefont
  {I.}~\bibnamefont {Ferrier-Barbut}}, \ and\ \bibinfo {author} {\bibfnamefont
  {T.}~\bibnamefont {Pfau}},\ }\href {\doibase 10.1038/nature16485} {\bibfield
  {journal} {\bibinfo  {journal} {Nature}\ }\textbf {\bibinfo {volume} {530}},\
  \bibinfo {pages} {194} (\bibinfo {year} {2016})}\BibitemShut {NoStop}%
\bibitem [{\citenamefont {Ferrier-Barbut}\ \emph
  {et~al.}(2016{\natexlab{b}})\citenamefont {Ferrier-Barbut}, \citenamefont
  {Schmitt}, \citenamefont {Wenzel}, \citenamefont {Kadau},\ and\ \citenamefont
  {Pfau}}]{ferrier_farbut_2016_2}%
  \BibitemOpen
  \bibfield  {author} {\bibinfo {author} {\bibfnamefont {I.}~\bibnamefont
  {Ferrier-Barbut}}, \bibinfo {author} {\bibfnamefont {M.}~\bibnamefont
  {Schmitt}}, \bibinfo {author} {\bibfnamefont {M.}~\bibnamefont {Wenzel}},
  \bibinfo {author} {\bibfnamefont {H.}~\bibnamefont {Kadau}}, \ and\ \bibinfo
  {author} {\bibfnamefont {T.}~\bibnamefont {Pfau}},\ }\href {\doibase
  10.1088/0953-4075/49/21/214004} {\bibfield  {journal} {\bibinfo  {journal}
  {Journal of Physics B: Atomic, Molecular and Optical Physics}\ }\textbf
  {\bibinfo {volume} {49}},\ \bibinfo {pages} {214004} (\bibinfo {year}
  {2016}{\natexlab{b}})}\BibitemShut {NoStop}%
\bibitem [{\citenamefont {Schmitt}\ \emph {et~al.}(2016)\citenamefont
  {Schmitt}, \citenamefont {Wenzel}, \citenamefont {B{\"o}ttcher},
  \citenamefont {Ferrier-Barbut},\ and\ \citenamefont {Pfau}}]{Schmitt2016}%
  \BibitemOpen
  \bibfield  {author} {\bibinfo {author} {\bibfnamefont {M.}~\bibnamefont
  {Schmitt}}, \bibinfo {author} {\bibfnamefont {M.}~\bibnamefont {Wenzel}},
  \bibinfo {author} {\bibfnamefont {F.}~\bibnamefont {B{\"o}ttcher}}, \bibinfo
  {author} {\bibfnamefont {I.}~\bibnamefont {Ferrier-Barbut}}, \ and\ \bibinfo
  {author} {\bibfnamefont {T.}~\bibnamefont {Pfau}},\ }\href {\doibase
  10.1038/nature20126} {\bibfield  {journal} {\bibinfo  {journal} {Nature}\
  }\textbf {\bibinfo {volume} {539}},\ \bibinfo {pages} {259} (\bibinfo {year}
  {2016})}\BibitemShut {NoStop}%
\bibitem [{\citenamefont {Tanzi}\ \emph {et~al.}(2019)\citenamefont {Tanzi},
  \citenamefont {Lucioni}, \citenamefont {Fam\`a}, \citenamefont {Catani},
  \citenamefont {Fioretti}, \citenamefont {Gabbanini}, \citenamefont {Bisset},
  \citenamefont {Santos},\ and\ \citenamefont {Modugno}}]{F1}%
  \BibitemOpen
  \bibfield  {author} {\bibinfo {author} {\bibfnamefont {L.}~\bibnamefont
  {Tanzi}}, \bibinfo {author} {\bibfnamefont {E.}~\bibnamefont {Lucioni}},
  \bibinfo {author} {\bibfnamefont {F.}~\bibnamefont {Fam\`a}}, \bibinfo
  {author} {\bibfnamefont {J.}~\bibnamefont {Catani}}, \bibinfo {author}
  {\bibfnamefont {A.}~\bibnamefont {Fioretti}}, \bibinfo {author}
  {\bibfnamefont {C.}~\bibnamefont {Gabbanini}}, \bibinfo {author}
  {\bibfnamefont {R.~N.}\ \bibnamefont {Bisset}}, \bibinfo {author}
  {\bibfnamefont {L.}~\bibnamefont {Santos}}, \ and\ \bibinfo {author}
  {\bibfnamefont {G.}~\bibnamefont {Modugno}},\ }\href {\doibase
  10.1103/PhysRevLett.122.130405} {\bibfield  {journal} {\bibinfo  {journal}
  {Phys. Rev. Lett.}\ }\textbf {\bibinfo {volume} {122}},\ \bibinfo {pages}
  {130405} (\bibinfo {year} {2019})}\BibitemShut {NoStop}%
\bibitem [{\citenamefont {Roccuzzo}\ \emph {et~al.}(2020)\citenamefont
  {Roccuzzo}, \citenamefont {Gallem\'{\i}}, \citenamefont {Recati},\ and\
  \citenamefont {Stringari}}]{roccuzzo2020}%
  \BibitemOpen
  \bibfield  {author} {\bibinfo {author} {\bibfnamefont {S.~M.}\ \bibnamefont
  {Roccuzzo}}, \bibinfo {author} {\bibfnamefont {A.}~\bibnamefont
  {Gallem\'{\i}}}, \bibinfo {author} {\bibfnamefont {A.}~\bibnamefont
  {Recati}}, \ and\ \bibinfo {author} {\bibfnamefont {S.}~\bibnamefont
  {Stringari}},\ }\href {\doibase 10.1103/PhysRevLett.124.045702} {\bibfield
  {journal} {\bibinfo  {journal} {Phys. Rev. Lett.}\ }\textbf {\bibinfo
  {volume} {124}},\ \bibinfo {pages} {045702} (\bibinfo {year}
  {2020})}\BibitemShut {NoStop}%
\bibitem [{\citenamefont {Chomaz}\ \emph {et~al.}(2019)\citenamefont {Chomaz},
  \citenamefont {Petter}, \citenamefont {Ilzh\"ofer}, \citenamefont {Natale},
  \citenamefont {Trautmann}, \citenamefont {Politi}, \citenamefont
  {Durastante}, \citenamefont {van Bijnen}, \citenamefont {Patscheider},
  \citenamefont {Sohmen}, \citenamefont {Mark},\ and\ \citenamefont
  {Ferlaino}}]{I1}%
  \BibitemOpen
  \bibfield  {author} {\bibinfo {author} {\bibfnamefont {L.}~\bibnamefont
  {Chomaz}}, \bibinfo {author} {\bibfnamefont {D.}~\bibnamefont {Petter}},
  \bibinfo {author} {\bibfnamefont {P.}~\bibnamefont {Ilzh\"ofer}}, \bibinfo
  {author} {\bibfnamefont {G.}~\bibnamefont {Natale}}, \bibinfo {author}
  {\bibfnamefont {A.}~\bibnamefont {Trautmann}}, \bibinfo {author}
  {\bibfnamefont {C.}~\bibnamefont {Politi}}, \bibinfo {author} {\bibfnamefont
  {G.}~\bibnamefont {Durastante}}, \bibinfo {author} {\bibfnamefont {R.~M.~W.}\
  \bibnamefont {van Bijnen}}, \bibinfo {author} {\bibfnamefont
  {A.}~\bibnamefont {Patscheider}}, \bibinfo {author} {\bibfnamefont
  {M.}~\bibnamefont {Sohmen}}, \bibinfo {author} {\bibfnamefont {M.~J.}\
  \bibnamefont {Mark}}, \ and\ \bibinfo {author} {\bibfnamefont
  {F.}~\bibnamefont {Ferlaino}},\ }\href {\doibase 10.1103/PhysRevX.9.021012}
  {\bibfield  {journal} {\bibinfo  {journal} {Phys. Rev. X}\ }\textbf {\bibinfo
  {volume} {9}},\ \bibinfo {pages} {021012} (\bibinfo {year}
  {2019})}\BibitemShut {NoStop}%
\bibitem [{\citenamefont {Natale}\ \emph {et~al.}(2019)\citenamefont {Natale},
  \citenamefont {van Bijnen}, \citenamefont {Patscheider}, \citenamefont
  {Petter}, \citenamefont {Mark}, \citenamefont {Chomaz},\ and\ \citenamefont
  {Ferlaino}}]{I2}%
  \BibitemOpen
  \bibfield  {author} {\bibinfo {author} {\bibfnamefont {G.}~\bibnamefont
  {Natale}}, \bibinfo {author} {\bibfnamefont {R.~M.~W.}\ \bibnamefont {van
  Bijnen}}, \bibinfo {author} {\bibfnamefont {A.}~\bibnamefont {Patscheider}},
  \bibinfo {author} {\bibfnamefont {D.}~\bibnamefont {Petter}}, \bibinfo
  {author} {\bibfnamefont {M.~J.}\ \bibnamefont {Mark}}, \bibinfo {author}
  {\bibfnamefont {L.}~\bibnamefont {Chomaz}}, \ and\ \bibinfo {author}
  {\bibfnamefont {F.}~\bibnamefont {Ferlaino}},\ }\href {\doibase
  10.1103/PhysRevLett.123.050402} {\bibfield  {journal} {\bibinfo  {journal}
  {Phys. Rev. Lett.}\ }\textbf {\bibinfo {volume} {123}},\ \bibinfo {pages}
  {050402} (\bibinfo {year} {2019})}\BibitemShut {NoStop}%
\bibitem [{\citenamefont {B\"ottcher}\ \emph {et~al.}(2019)\citenamefont
  {B\"ottcher}, \citenamefont {Schmidt}, \citenamefont {Wenzel}, \citenamefont
  {Hertkorn}, \citenamefont {Guo}, \citenamefont {Langen},\ and\ \citenamefont
  {Pfau}}]{S1}%
  \BibitemOpen
  \bibfield  {author} {\bibinfo {author} {\bibfnamefont {F.}~\bibnamefont
  {B\"ottcher}}, \bibinfo {author} {\bibfnamefont {J.-N.}\ \bibnamefont
  {Schmidt}}, \bibinfo {author} {\bibfnamefont {M.}~\bibnamefont {Wenzel}},
  \bibinfo {author} {\bibfnamefont {J.}~\bibnamefont {Hertkorn}}, \bibinfo
  {author} {\bibfnamefont {M.}~\bibnamefont {Guo}}, \bibinfo {author}
  {\bibfnamefont {T.}~\bibnamefont {Langen}}, \ and\ \bibinfo {author}
  {\bibfnamefont {T.}~\bibnamefont {Pfau}},\ }\href {\doibase
  10.1103/PhysRevX.9.011051} {\bibfield  {journal} {\bibinfo  {journal} {Phys.
  Rev. X}\ }\textbf {\bibinfo {volume} {9}},\ \bibinfo {pages} {011051}
  (\bibinfo {year} {2019})}\BibitemShut {NoStop}%
\bibitem [{\citenamefont {Guo}\ \emph {et~al.}(2019)\citenamefont {Guo},
  \citenamefont {B{\"{o}}ttcher}, \citenamefont {Hertkorn}, \citenamefont
  {Schmidt}, \citenamefont {Wenzel}, \citenamefont {B{\"{u}}chler},
  \citenamefont {Langen},\ and\ \citenamefont {Pfau}}]{S2}%
  \BibitemOpen
  \bibfield  {author} {\bibinfo {author} {\bibfnamefont {M.}~\bibnamefont
  {Guo}}, \bibinfo {author} {\bibfnamefont {F.}~\bibnamefont {B{\"{o}}ttcher}},
  \bibinfo {author} {\bibfnamefont {J.}~\bibnamefont {Hertkorn}}, \bibinfo
  {author} {\bibfnamefont {J.-N.}\ \bibnamefont {Schmidt}}, \bibinfo {author}
  {\bibfnamefont {M.}~\bibnamefont {Wenzel}}, \bibinfo {author} {\bibfnamefont
  {H.~P.}\ \bibnamefont {B{\"{u}}chler}}, \bibinfo {author} {\bibfnamefont
  {T.}~\bibnamefont {Langen}}, \ and\ \bibinfo {author} {\bibfnamefont
  {T.}~\bibnamefont {Pfau}},\ }\href {\doibase 10.1038/s41586-019-1569-5}
  {\bibfield  {journal} {\bibinfo  {journal} {Nature}\ }\textbf {\bibinfo
  {volume} {574}},\ \bibinfo {pages} {386} (\bibinfo {year}
  {2019})}\BibitemShut {NoStop}%
\bibitem [{\citenamefont {Lu}\ \emph {et~al.}(2010)\citenamefont {Lu},
  \citenamefont {Lu}, \citenamefont {Zhang}, \citenamefont {Qiu}, \citenamefont
  {Pu},\ and\ \citenamefont {Yi}}]{Lu2010}%
  \BibitemOpen
  \bibfield  {author} {\bibinfo {author} {\bibfnamefont {H.-Y.}\ \bibnamefont
  {Lu}}, \bibinfo {author} {\bibfnamefont {H.}~\bibnamefont {Lu}}, \bibinfo
  {author} {\bibfnamefont {J.-N.}\ \bibnamefont {Zhang}}, \bibinfo {author}
  {\bibfnamefont {R.-Z.}\ \bibnamefont {Qiu}}, \bibinfo {author} {\bibfnamefont
  {H.}~\bibnamefont {Pu}}, \ and\ \bibinfo {author} {\bibfnamefont
  {S.}~\bibnamefont {Yi}},\ }\href {\doibase 10.1103/PhysRevA.82.023622}
  {\bibfield  {journal} {\bibinfo  {journal} {Phys. Rev. A}\ }\textbf {\bibinfo
  {volume} {82}},\ \bibinfo {pages} {023622} (\bibinfo {year}
  {2010})}\BibitemShut {NoStop}%
\bibitem [{\citenamefont {Zhang}\ \emph {et~al.}(2019)\citenamefont {Zhang},
  \citenamefont {Maucher},\ and\ \citenamefont {Pohl}}]{Pohl2019}%
  \BibitemOpen
  \bibfield  {author} {\bibinfo {author} {\bibfnamefont {Y.-C.}\ \bibnamefont
  {Zhang}}, \bibinfo {author} {\bibfnamefont {F.}~\bibnamefont {Maucher}}, \
  and\ \bibinfo {author} {\bibfnamefont {T.}~\bibnamefont {Pohl}},\ }\href
  {\doibase 10.1103/PhysRevLett.123.015301} {\bibfield  {journal} {\bibinfo
  {journal} {Phys. Rev. Lett.}\ }\textbf {\bibinfo {volume} {123}},\ \bibinfo
  {pages} {015301} (\bibinfo {year} {2019})}\BibitemShut {NoStop}%
\bibitem [{\citenamefont {Lima}\ and\ \citenamefont {Pelster}(2012)}]{Pelster}%
  \BibitemOpen
  \bibfield  {author} {\bibinfo {author} {\bibfnamefont {A.~R.~P.}\
  \bibnamefont {Lima}}\ and\ \bibinfo {author} {\bibfnamefont {A.}~\bibnamefont
  {Pelster}},\ }\href {\doibase 10.1103/PhysRevA.86.063609} {\bibfield
  {journal} {\bibinfo  {journal} {Phys. Rev. A}\ }\textbf {\bibinfo {volume}
  {86}},\ \bibinfo {pages} {063609} (\bibinfo {year} {2012})}\BibitemShut
  {NoStop}%
\bibitem [{\citenamefont {W\"achtler}\ and\ \citenamefont
  {Santos}(2016)}]{Wachtler2016}%
  \BibitemOpen
  \bibfield  {author} {\bibinfo {author} {\bibfnamefont {F.}~\bibnamefont
  {W\"achtler}}\ and\ \bibinfo {author} {\bibfnamefont {L.}~\bibnamefont
  {Santos}},\ }\href {\doibase 10.1103/PhysRevA.93.061603} {\bibfield
  {journal} {\bibinfo  {journal} {Phys. Rev. A}\ }\textbf {\bibinfo {volume}
  {93}},\ \bibinfo {pages} {061603} (\bibinfo {year} {2016})}\BibitemShut
  {NoStop}%
\bibitem [{\citenamefont {Gallem\'{\i}}\ \emph {et~al.}(2020)\citenamefont
  {Gallem\'{\i}}, \citenamefont {Roccuzzo}, \citenamefont {Stringari},\ and\
  \citenamefont {Recati}}]{gallemi2020}%
  \BibitemOpen
  \bibfield  {author} {\bibinfo {author} {\bibfnamefont {A.}~\bibnamefont
  {Gallem\'{\i}}}, \bibinfo {author} {\bibfnamefont {S.~M.}\ \bibnamefont
  {Roccuzzo}}, \bibinfo {author} {\bibfnamefont {S.}~\bibnamefont {Stringari}},
  \ and\ \bibinfo {author} {\bibfnamefont {A.}~\bibnamefont {Recati}},\ }\href
  {\doibase 10.1103/PhysRevA.102.023322} {\bibfield  {journal} {\bibinfo
  {journal} {Phys. Rev. A}\ }\textbf {\bibinfo {volume} {102}},\ \bibinfo
  {pages} {023322} (\bibinfo {year} {2020})}\BibitemShut {NoStop}%
\bibitem [{\citenamefont {Hertkorn}\ \emph {et~al.}(2021)\citenamefont
  {Hertkorn}, \citenamefont {Schmidt}, \citenamefont {Guo}, \citenamefont
  {Böttcher}, \citenamefont {Ng}, \citenamefont {Graham}, \citenamefont
  {Uerlings}, \citenamefont {Langen}, \citenamefont {Zwierlein},\ and\
  \citenamefont {Pfau}}]{hertkorn2021}%
  \BibitemOpen
  \bibfield  {author} {\bibinfo {author} {\bibfnamefont {J.}~\bibnamefont
  {Hertkorn}}, \bibinfo {author} {\bibfnamefont {J.~N.}\ \bibnamefont
  {Schmidt}}, \bibinfo {author} {\bibfnamefont {M.}~\bibnamefont {Guo}},
  \bibinfo {author} {\bibfnamefont {F.}~\bibnamefont {Böttcher}}, \bibinfo
  {author} {\bibfnamefont {K.~S.~H.}\ \bibnamefont {Ng}}, \bibinfo {author}
  {\bibfnamefont {S.~D.}\ \bibnamefont {Graham}}, \bibinfo {author}
  {\bibfnamefont {P.}~\bibnamefont {Uerlings}}, \bibinfo {author}
  {\bibfnamefont {T.}~\bibnamefont {Langen}}, \bibinfo {author} {\bibfnamefont
  {M.}~\bibnamefont {Zwierlein}}, \ and\ \bibinfo {author} {\bibfnamefont
  {T.}~\bibnamefont {Pfau}},\ }\href@noop {} {\enquote {\bibinfo {title}
  {Pattern formation in quantum ferrofluids: from supersolids to
  superglasses},}\ } (\bibinfo {year} {2021}),\ \Eprint
  {http://arxiv.org/abs/2103.13930} {arXiv:2103.13930 [cond-mat.quant-gas]}
  \BibitemShut {NoStop}%
\bibitem [{\citenamefont {Zhang}\ \emph {et~al.}(2021)\citenamefont {Zhang},
  \citenamefont {Pohl},\ and\ \citenamefont {Maucher}}]{zhang2021}%
  \BibitemOpen
  \bibfield  {author} {\bibinfo {author} {\bibfnamefont {Y.-C.}\ \bibnamefont
  {Zhang}}, \bibinfo {author} {\bibfnamefont {T.}~\bibnamefont {Pohl}}, \ and\
  \bibinfo {author} {\bibfnamefont {F.}~\bibnamefont {Maucher}},\ }\href@noop
  {} {\enquote {\bibinfo {title} {Phases of supersolids in confined dipolar
  bose-einstein condensates},}\ } (\bibinfo {year} {2021}),\ \Eprint
  {http://arxiv.org/abs/2103.12688} {arXiv:2103.12688 [cond-mat.quant-gas]}
  \BibitemShut {NoStop}%
\bibitem [{\citenamefont {Leggett}(1970)}]{Leggett1970}%
  \BibitemOpen
  \bibfield  {author} {\bibinfo {author} {\bibfnamefont {A.~J.}\ \bibnamefont
  {Leggett}},\ }\href {\doibase 10.1103/PhysRevLett.25.1543} {\bibfield
  {journal} {\bibinfo  {journal} {Phys. Rev. Lett.}\ }\textbf {\bibinfo
  {volume} {25}},\ \bibinfo {pages} {1543} (\bibinfo {year}
  {1970})}\BibitemShut {NoStop}%
\bibitem [{\citenamefont {Leggett}(1998)}]{Leggett1998}%
  \BibitemOpen
  \bibfield  {author} {\bibinfo {author} {\bibfnamefont {A.~J.}\ \bibnamefont
  {Leggett}},\ }\href {\doibase 10.1023/B:JOSS.0000033170.38619.6c} {\bibfield
  {journal} {\bibinfo  {journal} {Journal of Statistical Physics}\ }\textbf
  {\bibinfo {volume} {93}},\ \bibinfo {pages} {927} (\bibinfo {year}
  {1998})}\BibitemShut {NoStop}%
\bibitem [{Note1()}]{Note1}%
  \BibitemOpen
  \bibinfo {note} {S. Roccuzzo, PhD Thesis, in preparation}\BibitemShut
  {NoStop}%
\bibitem [{\citenamefont {Martone}\ \emph {et~al.}(2021)\citenamefont
  {Martone}, \citenamefont {Recati},\ and\ \citenamefont
  {Pavloff}}]{Martone2021}%
  \BibitemOpen
  \bibfield  {author} {\bibinfo {author} {\bibfnamefont {G.~I.}\ \bibnamefont
  {Martone}}, \bibinfo {author} {\bibfnamefont {A.}~\bibnamefont {Recati}}, \
  and\ \bibinfo {author} {\bibfnamefont {N.}~\bibnamefont {Pavloff}},\ }\href
  {\doibase 10.1103/PhysRevResearch.3.013143} {\bibfield  {journal} {\bibinfo
  {journal} {Phys. Rev. Research}\ }\textbf {\bibinfo {volume} {3}},\ \bibinfo
  {pages} {013143} (\bibinfo {year} {2021})}\BibitemShut {NoStop}%
\bibitem [{\citenamefont {Blakie}\ \emph {et~al.}(2020)\citenamefont {Blakie},
  \citenamefont {Baillie}, \citenamefont {Chomaz},\ and\ \citenamefont
  {Ferlaino}}]{Blakie_1d_1}%
  \BibitemOpen
  \bibfield  {author} {\bibinfo {author} {\bibfnamefont {P.~B.}\ \bibnamefont
  {Blakie}}, \bibinfo {author} {\bibfnamefont {D.}~\bibnamefont {Baillie}},
  \bibinfo {author} {\bibfnamefont {L.}~\bibnamefont {Chomaz}}, \ and\ \bibinfo
  {author} {\bibfnamefont {F.}~\bibnamefont {Ferlaino}},\ }\href {\doibase
  10.1103/PhysRevResearch.2.043318} {\bibfield  {journal} {\bibinfo  {journal}
  {Phys. Rev. Research}\ }\textbf {\bibinfo {volume} {2}},\ \bibinfo {pages}
  {043318} (\bibinfo {year} {2020})}\BibitemShut {NoStop}%
\bibitem [{\citenamefont {Ancilotto}\ \emph {et~al.}(2013)\citenamefont
  {Ancilotto}, \citenamefont {Rossi},\ and\ \citenamefont
  {Toigo}}]{Ancilotto2013}%
  \BibitemOpen
  \bibfield  {author} {\bibinfo {author} {\bibfnamefont {F.}~\bibnamefont
  {Ancilotto}}, \bibinfo {author} {\bibfnamefont {M.}~\bibnamefont {Rossi}}, \
  and\ \bibinfo {author} {\bibfnamefont {F.}~\bibnamefont {Toigo}},\ }\href
  {\doibase 10.1103/PhysRevA.88.033618} {\bibfield  {journal} {\bibinfo
  {journal} {Phys. Rev. A}\ }\textbf {\bibinfo {volume} {88}},\ \bibinfo
  {pages} {033618} (\bibinfo {year} {2013})}\BibitemShut {NoStop}%
\bibitem [{\citenamefont {Baillie}\ \emph {et~al.}(2017)\citenamefont
  {Baillie}, \citenamefont {Wilson},\ and\ \citenamefont
  {Blakie}}]{Blakie_droplet_bdg}%
  \BibitemOpen
  \bibfield  {author} {\bibinfo {author} {\bibfnamefont {D.}~\bibnamefont
  {Baillie}}, \bibinfo {author} {\bibfnamefont {R.~M.}\ \bibnamefont {Wilson}},
  \ and\ \bibinfo {author} {\bibfnamefont {P.~B.}\ \bibnamefont {Blakie}},\
  }\href {\doibase 10.1103/PhysRevLett.119.255302} {\bibfield  {journal}
  {\bibinfo  {journal} {Phys. Rev. Lett.}\ }\textbf {\bibinfo {volume} {119}},\
  \bibinfo {pages} {255302} (\bibinfo {year} {2017})}\BibitemShut {NoStop}%
\bibitem [{\citenamefont {Abad}\ \emph {et~al.}(2009)\citenamefont {Abad},
  \citenamefont {Guilleumas}, \citenamefont {Mayol}, \citenamefont {Pi},\ and\
  \citenamefont {Jezek}}]{Abad2009}%
  \BibitemOpen
  \bibfield  {author} {\bibinfo {author} {\bibfnamefont {M.}~\bibnamefont
  {Abad}}, \bibinfo {author} {\bibfnamefont {M.}~\bibnamefont {Guilleumas}},
  \bibinfo {author} {\bibfnamefont {R.}~\bibnamefont {Mayol}}, \bibinfo
  {author} {\bibfnamefont {M.}~\bibnamefont {Pi}}, \ and\ \bibinfo {author}
  {\bibfnamefont {D.~M.}\ \bibnamefont {Jezek}},\ }\href {\doibase
  10.1103/PhysRevA.79.063622} {\bibfield  {journal} {\bibinfo  {journal} {Phys.
  Rev. A}\ }\textbf {\bibinfo {volume} {79}},\ \bibinfo {pages} {063622}
  (\bibinfo {year} {2009})}\BibitemShut {NoStop}%
\bibitem [{\citenamefont {Regge}(1972)}]{regge_1972}%
  \BibitemOpen
  \bibfield  {author} {\bibinfo {author} {\bibfnamefont {T.}~\bibnamefont
  {Regge}},\ }\href {\doibase 10.1007/bf00655491} {\bibfield  {journal}
  {\bibinfo  {journal} {Journal of Low Temperature Physics}\ }\textbf {\bibinfo
  {volume} {9}},\ \bibinfo {pages} {123–133} (\bibinfo {year}
  {1972})}\BibitemShut {NoStop}%
\bibitem [{\citenamefont {Dalfovo}(1992)}]{Dalfovo91}%
  \BibitemOpen
  \bibfield  {author} {\bibinfo {author} {\bibfnamefont {F.}~\bibnamefont
  {Dalfovo}},\ }\href {\doibase 10.1103/PhysRevB.46.5482} {\bibfield  {journal}
  {\bibinfo  {journal} {Phys. Rev. B}\ }\textbf {\bibinfo {volume} {46}},\
  \bibinfo {pages} {5482} (\bibinfo {year} {1992})}\BibitemShut {NoStop}%
\bibitem [{\citenamefont {Chester}\ \emph {et~al.}(1968)\citenamefont
  {Chester}, \citenamefont {Metz},\ and\ \citenamefont {Reatto}}]{Chester68}%
  \BibitemOpen
  \bibfield  {author} {\bibinfo {author} {\bibfnamefont {G.~V.}\ \bibnamefont
  {Chester}}, \bibinfo {author} {\bibfnamefont {R.}~\bibnamefont {Metz}}, \
  and\ \bibinfo {author} {\bibfnamefont {L.}~\bibnamefont {Reatto}},\ }\href
  {\doibase 10.1103/PhysRev.175.275} {\bibfield  {journal} {\bibinfo  {journal}
  {Phys. Rev.}\ }\textbf {\bibinfo {volume} {175}},\ \bibinfo {pages} {275}
  (\bibinfo {year} {1968})}\BibitemShut {NoStop}%
\end{thebibliography}%


\clearpage
\appendix

\renewcommand\thefigure{\thesection S\arabic{figure}}   
\setcounter{figure}{0}   

\section{Supplemental Material}

\subsection{A. Quasi one-dimensional dipolar Bose gas in a box}

\begin{figure}
    \includegraphics[width=0.5\textwidth]{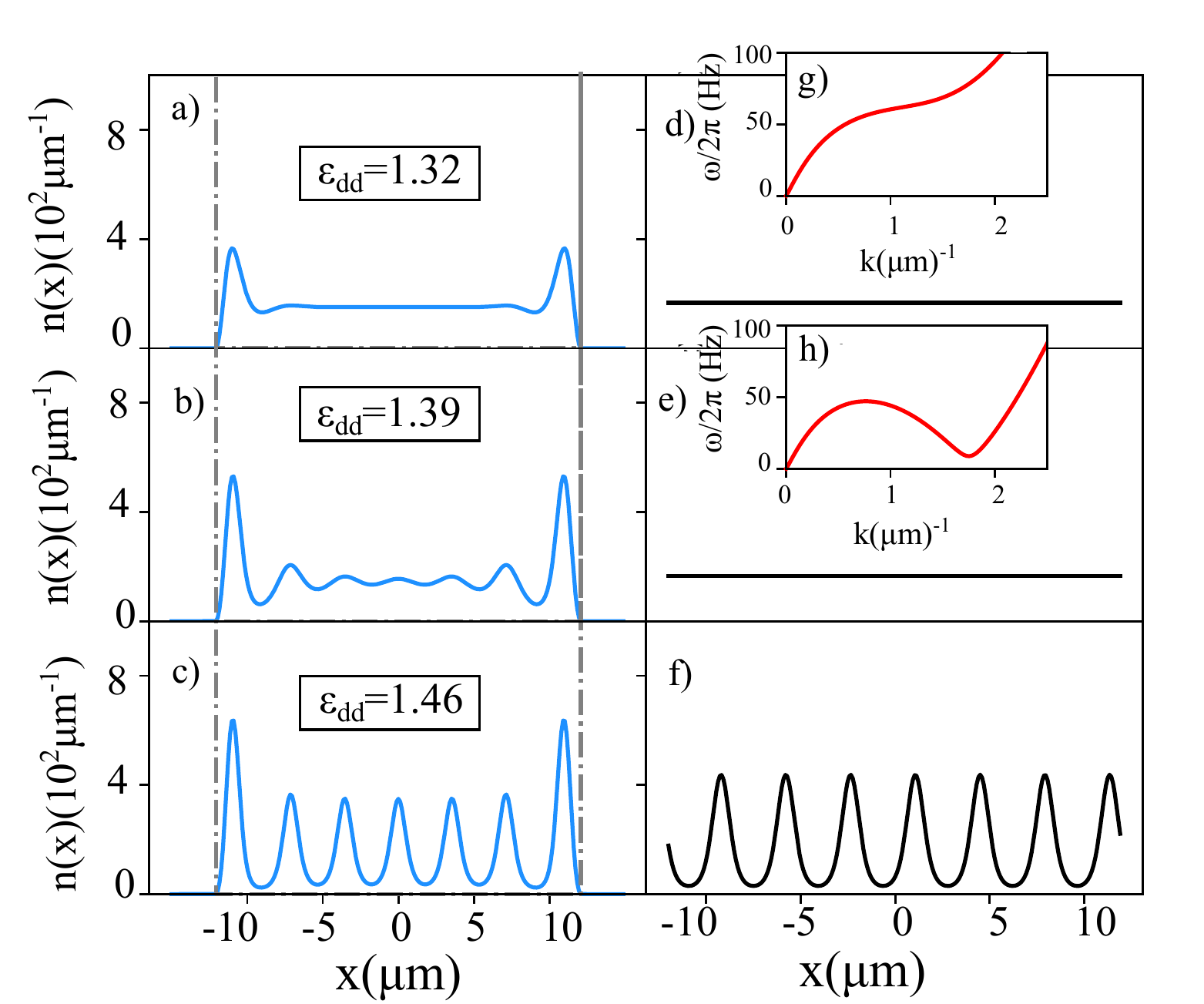}
    \caption{Ground state integrated density profiles
            $\mbox{n(x)=}\int\mbox{dydz}|\Psi\mbox{(x,y,z)}|^2$ of N=$4\times10^4$ atoms of
            $^{164}$Dy in a transverse harmonic confinement of frequencies
            $\omega_y=\omega_z=(2\pi)100$Hz confined by a box potential of height
            $\mbox{V}_0=100\hbar\omega_z$ at positions x=$\pm$12$\mu$m (panels a,b,c) or with
            periodic boundary conditions at x=$\pm$12$\mu$m (panels d,e,f). Insets g and h show the
            excitation spectrum calculated by solving the Bogolyubov-de Gennes equations for the
            configurations of panels d,e. The profiles reported in panels a,d (respectively, b,e and
            c,f) are calculated by fixing $\epsilon_{dd}$=1.32 (respectively, 1.39 and 1.46)}
    \label{fig_1d}
\end{figure}

 To complement our analysis of the behavior of a dipolar Bose gas in presence of hard walls, we have also considered the case of a quasi one-dimensional configuration confined by a box potential in the elongated direction. We have in particular determined the ground-state density profiles of $N=4\times10^4$ atoms of $^{164}$Dy confined  in
a trapping potential of the form $V_{ext}= \frac{1}{2}m(\omega^2_y y^2+\omega^2_z z^2) +
V_{box}(x,L)$, with $\omega_y=\omega_z=(2\pi)100$Hz, $V_{box}(x,L)=V_0$ for $|x|\geq L$ and 0
otherwise, $V_0 = 100\hbar\omega_z$ and L=12$\mu$m. Typical density profiles for different values of
$\epsilon_{dd}$ are reported in Fig. \ref{fig_1d} panels a, b and c.  
In the figure we also report
the density profiles calculated in a one-dimensional configuration, imposing periodic boundary conditions at $x=\pm$L (panels d,e,f) as well as the
corresponding excitation spectra (insets g,h) calculated in the uniform phase by solving the
Bogolyubov-de Gennes (BdG) equations. The latter are obtained by linearizing the eGPE around the
ground states (details on the procedure for dipolar gases described by the eGPE, can be
found in references \cite{Roccuzzo1,Ancilotto2013,Blakie_droplet_bdg}). The configurations d) and e) correspond to a superfluid phase. The configuration e) is
characterised by a pronounced roton minimum, precursor of the instability to a periodically
modulated density (supersolid phase) for larger values of $\epsilon_{dd}$ (panel f)). In the
presence of the box  atoms   accumulate close to the walls even for  small values of $\epsilon_{dd}$
(weakly interacting dipolar case, panels a,d and inset g), when the excitation spectrum of the uniform phase does not show a roton minimum. 
Due to the long range and anisotropic nature of the dipolar force, even in this case the density profile deeply differs from the  results holding for a one dimensional BEC interacting with a short range potential. In the latter case the density profile, near a hard wall located at $x=0$, is fixed by the healing length $\xi= \sqrt{\hbar^2/2mg\overline{n}}$ according to $n(x)= \overline{n}\tanh^2(x/\sqrt2 \xi)$ where $\overline{n}$ is the bulk density away from the edge of the box \cite{BecBook2016}.  The concept of healing length is not easily applicable to the case of a dipolar gas, whose different behavior  is due to the long-range nature of the force, the repulsive effect felt by the aligned dipoles, which tend to accumulate near the border, the presence, for large values of $\epsilon_{dd}$, of rotonic oscillations and, of course, the emergence of spontaneous density modulations characterizing the supersolid and the crystal phases. The emergence of the rotonic oscillations is reminiscent of a similar effect characterizing the  density profile in the vicinity of a quantized vortex \cite{Abad2009}. This effect, originally theoretically investigated for quantized
vortices in superfluid helium, is a direct consequence of the presence of the roton in the excitation spectrum \cite{regge_1972,Dalfovo91,Chester68}. A similar behavior is observed along each edge of the polygonal box potentials shown in figures \ref{square_box} and \ref{geometries} of the main text, suggesting that such edge configurations host, between two vertices of the confining potential, localized excitations corresponding to those that naturally occurs in quasi one-dimensional configurations.

\subsection{B. Roton energy for a quasi one-dimensional dipolar gas with periodic boundary conditions}
\begin{figure}[b!]
    \includegraphics[width = 0.5\textwidth]{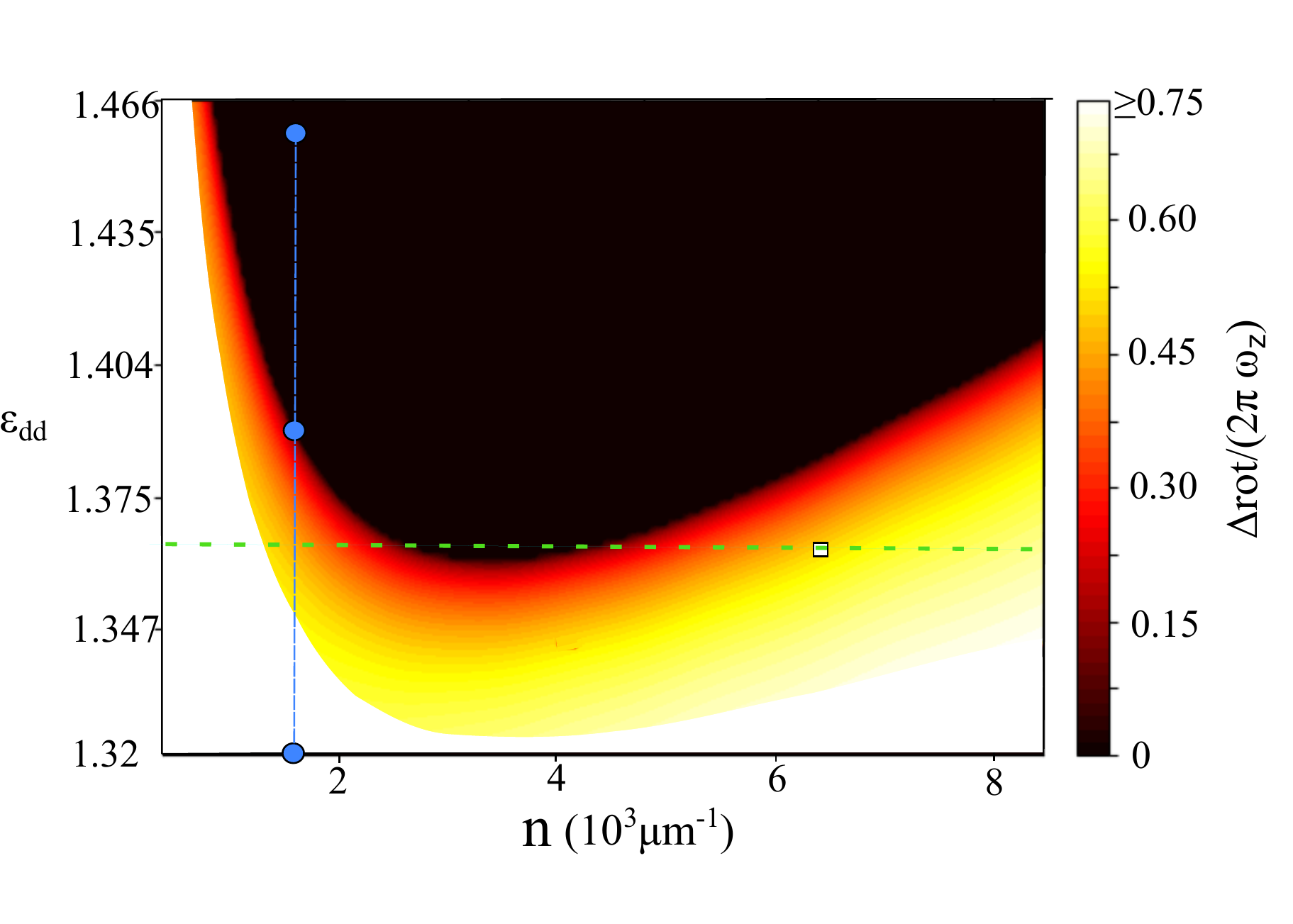}
    \caption{Roton energy of a quasi one-dimensional dipolar Bose gas with transverse harmonic confinement of frequencies $\omega_y=\omega_z=(2\pi)100$Hz, with
    periodic boundary conditions along the x axis, as function of the linear density $n=N/L$, where N is the number of atoms and L the length of the simulation cell along x, and of $\epsilon_{dd}$. The blue dots correspond, from bottom to top, to the configurations of figure \ref{fig_1d} panels d,e,f, while the white empty square corresponds to the the ring edge of the configuration shown in figure \ref{geometries} panel e (circular transverse confinement). The dark area corresponds to configurations in which the roton mode, calculated starting from a uniform configuration, is unstable and where spontaneous density modulations of supersolid or crystal nature are formed. The green, horizontal dashed line is a guide to the eye showing that, for fixed $\epsilon_{dd}$, starting from a low-density syperfluid, increasing the density results in a phase transition to a supersolid (and eventually to a droplet crystal), while at even higher densities the system turns superfluid again.} 
    \label{phase_diagram}
\end{figure}

In order to  understand the  (almost) uniform density phase appearing along the edges of the box configurations shown in figure \ref{geometries}, we have extended the results presented in \cite{Blakie_1d_1,hertkorn2021} by studying the value of the roton excitation energy calculated for a quasi one-dimensional uniform dipolar gas in a transverse harmonic confinement, with periodic boundary conditions along the unconfined direction, as a function of the linear density n = N/L, where n is the number of atoms and L the length of the tube, and of $\epsilon_{dd}$. The results are shown in figure \ref{phase_diagram}. Notice that, for a fixed value of $\epsilon_{dd}$ above a critical threshold (here, $\simeq 1.35$), increasing the density starting from  small values (see the green dashed line), the system undergoes a transition from a superfluid to a supersolid phase, revealed  by the instability of the roton miminum,  to come back  to the superfluid phase, characterized again by the occurrence of roton with a finite excitation energy, at  larger values of the density.

\end{document}